\documentclass[twocolumn,showpacs,preprintnumbers,prd,superscriptaddress,nofootinbib]{revtex4-1}
\usepackage{amsmath}
\usepackage{amsfonts}
\usepackage{amssymb}
\usepackage{graphicx}
\usepackage{color}
\usepackage{hyperref}
\usepackage{booktabs}

\def\eis{\emph{eis }}
\def\Eis{\emph{Eis }}

\AtBeginDocument{
\heavyrulewidth=.08em
\lightrulewidth=.05em
\cmidrulewidth=.03em
\belowrulesep=.65ex
\belowbottomsep=0pt
\aboverulesep=.4ex
\abovetopsep=0pt
\cmidrulesep=\doublerulesep
\cmidrulekern=.5em
\defaultaddspace=.5em
}

\begin{document}

\title{Toward unbiased estimations of the statefinder parameters}

%%%%%%%%%%%%%%%%%%%%%%%%%%%%%%%%%%%%%%%%%%%%%%%%%%%%%%%%%%%%%%

\author{Alejandro Aviles}
\email{avilescervantes@gmail.com}
\affiliation{
ABACUS-Centro de Matem\'aticas Aplicadas y C\'omputo de Alto Rendimiento,
Departamento de Matem\'aticas, Centro de Investigaci\'on y de Estudios Avanzados (Cinvestav-IPN),
Carretera M\'exico-Toluca km. 38.5, La Marquesa, 52740 Ocoyoacac,
Estado de M\'exico, Mexico.}
\affiliation{
Consejo Nacional de Ciencia y Tecnologı\'ia, Av. Insurgentes Sur 1582,
Colonia Cr\'edito Constructor, Del. Benito Jurez, 03940, Ciudad de M\'exico, M\'exico}
\affiliation{
Departamento de F\'isica, Instituto Nacional de Investigaciones Nucleares (ININ),
Carretera M\'exico-Toluca km. 36.5, La Marquesa, 52750 Ocoyoacac, Estado de M\'exico,
Mexico.}

\author{Jaime Klapp}
\email{jaime.klapp@inin.gob.mx}
\affiliation{
ABACUS-Centro de Matem\'aticas Aplicadas y C\'omputo de Alto Rendimiento,
Departamento de Matem\'aticas, Centro de Investigaci\'on y de Estudios Avanzados (Cinvestav-IPN),
Carretera M\'exico-Toluca km. 38.5, La Marquesa, 52740 Ocoyoacac,
Estado de M\'exico, Mexico.}
\affiliation{
Departamento de F\'isica, Instituto Nacional de Investigaciones Nucleares (ININ),
Carretera M\'exico-Toluca km. 36.5, La Marquesa, 52750 Ocoyoacac, Estado de M\'exico,
Mexico.}

\author{Orlando Luongo}
\email{luongo@na.infn.it}
\affiliation{Department of Mathematics and Applied Mathematics, University of Cape Town, Rondebosch 7701, Cape Town, South Africa.}
\affiliation{Astrophysics, Cosmology and Gravity Centre (ACGC), University of Cape Town, Rondebosch 7701, Cape Town, South Africa.}
\affiliation{Dipartimento di Fisica, Universit\`a di Napoli ''Federico II'', Via Cinthia, I-80126, Napoli, Italy.}
\affiliation{Istituto Nazionale di Fisica Nucleare (INFN), Sezione di Napoli, Via Cinthia, I-80126 Napoli, Italy.}

\begin{abstract}
With the use of simulated supernova catalogs, we show that the statefinder parameters turn out to be poorly and biased estimated by standard cosmography.
To this end, we compute their standard deviations and several bias statistics on cosmologies near the concordance model,
demonstrating that these are very large, making standard cosmography unsuitable for future and wider compilations of data.
To overcome this issue, we propose a new method that consists in introducing
the series of the Hubble function into the luminosity distance, instead of
considering the usual direct Taylor expansions of the luminosity distance.
Moreover, in order to speed up the numerical computations, we estimate the coefficients of our expansions
in a hierarchical manner, in which the order of the expansion depends on the redshift of every single piece of data.
In addition, we propose two hybrids methods that incorporates standard cosmography at low
redshifts. The methods presented here perform better
than the standard approach of cosmography both in the errors and bias of the estimated statefinders.
We further propose a one-parameter diagnostic to reject non-viable methods in cosmography.
\end{abstract}

\maketitle

\section{Introduction}

The universe is currently undergoing a late-time speeding up phase \cite{Riess:1998cb,Perlmutter:1998np}.
The component responsible for the
acceleration is commonly named \emph{dark energy} and manifests a negative equation
of state providing gravitational repulsive effects. Nonetheless, a totally satisfactory
fundamental description does not exist and dark energy still continues
challenging the concordance paradigm \cite{Frieman:2008sn,Tsujikawa:2010sc}. In addition, to characterize the dark energy evolution,
one needs to postulate a model \emph{a priori}. This leads to numerical constraints on the free coefficients of the model which are
determined in a \emph{model dependent way}.
Hence, treatments which encapsulate different aspects of
cosmology without calling any specific model become
useful to understand whether dark energy evolves or not in time.

In the $\Lambda$CDM model, the accelerated
expansion of the Universe is driven by a cosmological constant, while the
clustering of matter at large scales is a consequence of the gravitational
self-attraction of a stress-free and dust-like collection of particles dubbed cold dark matter.
Although the $\Lambda$CDM model suffers from the cosmological constant and cosmic coincidence
problems \cite{Weinberg:1988cp,Bianchi:2010uw}, the cosmological constant is physically well-motivated as the
minimal modification to General Relativity consistent with general
covariance. Perhaps more intriguing is the nature of the dark matter sector: for example,
today observations are not able to tell if it is really cold, or it is warm.

The success of the $\Lambda$CDM model is much more appreciable in the
early universe (see, e.g., figures 1 and 3 in
\cite{Ade:2015xua}), but it is still not very well tested at late
times, where there is large room for evolving dark energy and several behaviors for the dark matter.\footnote{Other
possibilities are the existence of unified fluids for the whole dark
sector \cite{Hu:1998tj,Bento:2002ps}, that the laws of gravity are different at
large scales or at late-times \cite{Capozziello:2002rd,Carroll:2003wy}, or even that dark energy comes from as a byproduct of
quantum effects \cite{Martin:2012bt}.} Hence, approaches beyond the $\Lambda$CDM model are widely investigated by the community.

Among general model independent treatments, {\it cosmography--on the
background} (hereafter cosmography)
attempts to reconstruct the expansion history of the universe in a model independent way; see \cite{Dunsby:2015ers} for a recent review.
To do so, it usually takes into account a set of parameters,
derivatives of the scale factor, related to the statefinders  \cite{Sahni:2002fz,Alam:2003sc}  and
generically named
\emph{cosmographic series}.\footnote{In our paper we refer to the
cosmographic series with the name statefinders, which slightly differs from the introduced first in \cite{Sahni:2002fz}.
For details see Appendix \ref{app::formulas}.} Such a procedure enables us to consider the fewest
number of assumptions as possible. Indeed, for the $\Lambda$CDM model at the background level,
one needs three quantities only to address the late evolution: the Hubble rate $H$, the deceleration
parameter $q$ and its variation, related to the jerk parameter $j$. Furthermore, it is usually assumed a flat universe,
which reduce the parameters to $H$ and $q$.
The need to go beyond these quantities has increased recently in order to constrain additional parameters with higher accuracy.
This has led cosmologists to propose new diagnostics to investigate the corresponding phase spaces of
coefficients \cite{Sahni:2002fz,Alam:2003sc,Arabsalmani:2011fz}.

To reach the objective of measuring the
statefinders, several attempts have been considered. For
example: fits with standard cosmography (SC) \cite{Weinberg:100595,Sahni:2002fz,Alam:2003sc}, Pad\'e rational approximants
\cite{Gruber:2013wua,Aviles:2014rma,Zhou:2016nik}, expansions on different functions of
redshift $z$ \cite{Cattoen:2007sk,Aviles:2012ay},
principal component analysis \cite{Qin:2015eda,Feng:2016stb}, Gaussian
process cosmography \cite{Shafieloo:2012ht,Nair:2013sna}, and cubic spline reconstructions of the Hubble function \cite{Bernal:2016gxb}, among others.
However, beyond the first order statefinder term, $q$, none of these approaches turn out to be
totally satisfactory.

 The case of fitting theoretical models by using the statefinders has always attracted attention. More recently, it has been argued that
 the impossibility for the statefinders to constrain general models poses questions about
 its usefulness \cite{Busti:2015xqa,delaCruz-Dombriz:2016rxm,Saez-Gomez:2016vfa}.
 In this regard, our point of view is mostly different. Let us suppose to have a theoretical model
 $\mathcal{M}(\Psi; \alpha_i )$ with $\Psi$ denoting the fields of the model and $\alpha_i$ its free parameters. For $\mathcal{M}$ to be
 well defined, each realization of the parameters, subjected to initial conditions, should give a unique Hubble
 diagram\footnote{It is not always possible to obtain the Hubble function for arbitrary free parameters.
 Particularly in higher derivatives theories.};
 and given that, it is as simple as taking derivatives to find the statefinders in that model.
 Thereafter, one can make comparisons to the measured statefinders,
 and accept or reject the realization $\alpha_i$; clearly, one can restart the procedure with a new realization of the parameters,
 although at this point it could be a better idea to fit directly to the data. That is, to fit a general class of models,
 say {\it e.g.} the whole class of $f(R)$ theories,
 using the statefinders is not always possible at least some extra assumptions are
 considered \cite{Aviles:2012ir}.

 Although doable, the main objective of cosmography is not to constrain theoretical models ---the statefinders are not data.
 Instead, it is to reconstruct the Hubble diagram as model independent as possible. Cosmography is a ``top-down'' approach to
 cosmology \cite{Shafieloo:2012ht}, that
 attempts to deduce its kinematics directly from observations; contrary to a ``bottom-up'' approach,
 that assumes the dynamics of a given model from the very beginning.

In the class of cosmographic methods lying on series expansions, the most spinous difficulty remains likely the convergence problem \cite{Cattoen:2007sk}. It is
intimately related to truncating series choosing the particular Taylor expansion to use.
% One can imagine that simply expanding about $z=0$ and using those
% expansions with data could be predictive enough. Unfortunately, data span in regimes which overcome the condition $z\simeq
% 0$, giving rise to the following conceptual problem:
% how can we use Taylor expansions centered about $z=0$, matching those
% expansions with data exceeding those regimes?
Indeed, all finite Taylor series diverge when $z\gg 1$
leading to possible misleading outcomes. Some authors have attempted to overcome this obstacle by using auxiliary variables
built up in terms of the redshift $z$ \cite{Cattoen:2007sk,Capozziello:2011tj,Aviles:2012ay}.

Another major problem of the available cosmographical methods is that the
estimation of the statefinder parameters is in general
biased. For example, the authors of Ref.~\cite{Busti:2015xqa} conclude that
estimations in cosmography are biased when expanding the luminosity distance as powers of the function of redshift $y=z/(1+z)$.

\begin{figure}
\begin{center}
\includegraphics[width=3.2in]{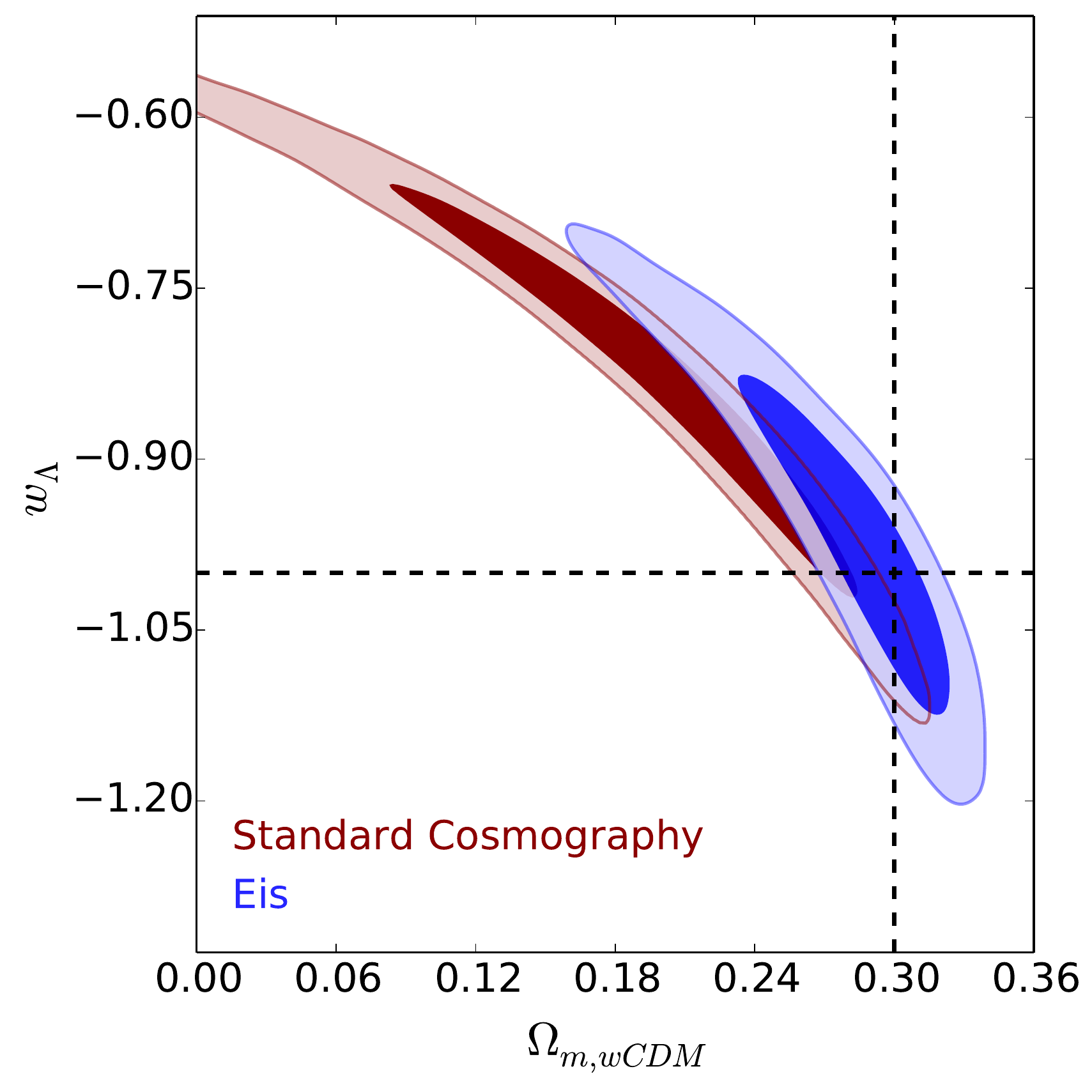}
\caption{2D confidence regions for derived parameters $\Omega_m$ and $w$ for both standard cosmography
and the method of this work (\emph{Eis}). Here we fit only the first two cosmographic parameters to our simulated exact data built upon
a $w$CDM model with $\Omega_m = 0.3$ and $w=-1$. We note that standard cosmography fails to obtain the true cosmology at $2\sigma$.}
\label{fig:01}
\end{center}
\end{figure}

SC also suffers from this bias problem. To neatly observe it, we consider the following simple example. In this approach, the
luminosity distance $d_L$  is given by
\begin{eqnarray} \label{dL_sc}
 & & \tilde{d}_{L \text{(SC)}}(z;q_0,j_0,s_0,\dots) = z + \frac{1}{2}
(1-q_0) z^2 \nonumber\\
 & & + \frac{1}{6}  (-1 + q_0 + 3 q_0^2  - j_0 )z^3 \nonumber\\
 & &  + \frac{1}{24}  \left( 2 + 5 j_0 -2 q_0 +10 j_0 q_0 - 15 q_0^2 - 15
q_0^3 + s_0 \right) z^4 \nonumber\\
 & &  + \cdots
\end{eqnarray}
where $\tilde{d_L} \equiv H_0 d_L / c$ with $c = 1$ the speed of light and
a subindex ``0'' means that the statefinders are evaluated at $z=0$.
We fit the first two cosmographic parameters in SC to our ``exact'' simulated
catalog with $\Omega_m=0.3$ and $w=-1$ (see Sec.~\ref{subsec:exactSD} below). From this fit we can derive the
parameters of a $w$CDM model. In Fig.~\ref{fig:01} we show a plot
of the joint 2-dimensional posterior region $\Omega_m$-$w$.
It can be observed that the
true cosmology is out of the estimations of SC at $2\sigma$, showing the large bias in SC.

Another issue in cosmography is due to correlations. In general, the statefinders are degenerated, for example in the flat $\Lambda$CDM model there is the
linear degeneracy $s_{0,\,\Lambda\text{CDM}} = -2 - 3 q_{0,\,\Lambda\text{CDM}}$. Since the \emph{true} cosmological
model should be  close to $\Lambda\text{CDM}$, we expect that
a successful cosmography method should follow this trend in a small neighborhood of its best fit.\footnote{We do not expect that the linear
degeneracy is followed exactly away from the best fit, even for the exact simulated data, because variations of the jerk parameter breaks it.}
In Sec.~\ref{sec:sdanalysis},
we show that this is not the case for SC.
We assume the reason behind it is that the third-degree polynomial form of the $z^4$ coefficient in Eq.~(\ref{dL_sc}) leads to
fictitious degeneracies.

In this work, we extensively use simulated supernova catalogs to address the bias problem and to study the posterior distributions in SC.
We show that the posteriors not only have wide dispersions, but also a large bias, non-tolerable for future observations (for example, [cite]).
To overcome these problems, and motivated for the unwanted degeneracies in SC exposed above,
we propose a new route that lies on expanding the Hubble function and use that expansion as an input for observables as the
modulus distance. By using several tests, we indeed show that the dispersions and
bias turn out to be smaller. We consider our approach by means of a hierarchy in which we split the
data into redshift domains, speeding up the numerical computations. Due to the success of SC at low redshifts we further propose two hybrid methods.

The paper is organized as follows:  in Sec.~\ref{cosmografia} we highlight the treatment of cosmography, whereas in Sec.~\ref{sec:model},
we present our proposed
cosmographic approach. In Sec.~\ref{sec:sdanalysis}, we go toward it and
we test our model and SC with simulated data, while
in Sec.~\ref{sec:rdanalysis} we estimate the statefinders parameters using
the JLA and Union2.1 compilations. Finally, in Sec.~\ref{sec:concl} we present our conclusions and
perspectives.

\begin{section}{Cosmography of the universe}
\label{cosmografia}

Cosmography, or better the \emph{cosmographic method}, stands for a model-independent treatment able to fix limits on universe's kinematics, 
without imposing a cosmological model \emph{a priori} \cite{Cattoen:2007sk,Aviles:2012ay}. Particularly, cosmography makes possible to gather 
the universe expansion history at small redshift regimes by simply invoking the cosmological principle, 
without any additional requirements on Einstein's energy momentum tensor. In other words, postulating the homogeneity and isotropy, 
with spatial curvature somehow fixed, cosmography can account for the evolving dark energy contributions in Einstein's equations. To do so, one 
can understand if dark energy is composed by a pure cosmological constant or by some particular exotic fluid.

The strategy is based on expanding all cosmological observables of interest around present time. Even though this procedure is the most used, it 
is even possible to expand only the scale factor $a(t)$ and then to rewrite all quantities in terms of it. In any cases, each expansions may be 
matched with cosmic data in order to give bounds on the evolution of each variable under exam. Hence, one gets numerical outcomes which do not 
depend on any particular requirements since only Taylor expansions are compared with data. Further, relating observations to theoretical 
predictions means to heal the degeneracy problem
among cosmological models. Indeed, cosmography is therefore able to distinguish different classes of models which turn out to be compatible 
with
cosmographic predictions and those ones that have to be discarded.

Bearing this in mind leads to consider cosmography as a powerful method that allows to study present-time cosmology and  to describe the dynamical 
evolution of the universe. Following standard recipes lies on postulating a scale factor $a\equiv1/(1+z)$ expanded today as:
\begin{equation}
a(t)=1+\sum_{k=1}^{\infty}\dfrac{1}{k!}\dfrac{d^k a}{dt^k}\bigg | _{t=t_0}(t-t_0)^k\,.
\label{eq:scale factor}
\end{equation}
The \textit{cosmographic parameters} are the terms entering the expansion \eqref{eq:scale factor}\footnote{They have been explicitly reported in Appendix A.}, 
which represent model-independent quantities related to the Taylor expansion of Hubble's rate by:
\begin{equation}
H(z)=H_0\left(1 + \sum_{\ell=1}^{\infty}\dfrac{1}{\ell!}\dfrac{d^\ell H}{dz^\ell}\bigg | _{z=0} z^\ell \right)\,.
\label{eq:Taylor H}
\end{equation}
This gives:
\begin{align}
H_z\big | _{z=0}& =1 + q_0\ , \nonumber\\
H_{zz}\big | _{z=0}&=j_0 - q_0^2\ ,\\
H_{zzz}\big | _{z=0}&=\dfrac{1}{6}\Big(j_0(3 +4q_0)-3q_0(1+q_0)+s_0\Big)\ .\nonumber
\end{align}
where we used the convention in which the subscripts represent the derivative orders with respect to the redshift $z$. In analogy, we can explicitly write Eq. \eqref{eq:scale factor} to have:
\begin{eqnarray}\label{serie1a2}
a(t)  & \sim & 1+  H_0 \Delta t - \frac{q_0}{2}  H_0^2\Delta t^2+\nonumber \\
&+&
\frac{j_0}{6} H_0^3 \Delta t^3 +   \frac{s_0}{24}  H_0^4\Delta t^4+\ldots\ ,
\end{eqnarray}
where we have normalized the scale factor to $a(t_0) = 1$.

Each term displays a specific meaning associated to dynamical properties of the universe. For example, $q$ and $j$ fix kinematic properties at lower redshift domains, since the  value of $q$ at a given time specifies whether the universe is accelerating or decelerating. Further $j$ is intimately related to its variation in the past. In particular,
\begin{enumerate}
\item
    $\underline{q_0>0}$ shows that the universe is expanding universe although it undergoes a deceleration phase, as for example in the case of pressureless matter dominated phase.

\item
    $\underline{-1<q_0<0}$ represents an expanding universe which speeds up as expected by current observations.

\item
    $\underline{q_0=-1}$ indicates that all the whole cosmological energy budget is dominated by a de Sitter fluid.

 \item   $\underline{j_0<0}$ means that dark energy influences early time dynamics in the same way of late time evolution.

  \item  $\underline{j_0=0}$ indicates that the acceleration parameter smoothly tends to a precise value, without changing
  its behavior as $z\rightarrow\infty$.

\item
 $\underline{j_0>0}$ implies that the universe acceleration started at a precise time during the  evolution, associated to the transition redshift. 
 In such a way, it provides the acceleration changes sign during time.

\end{enumerate}

All those properties are valid since the variation of $q$ and $j$ are intertwined by the following relation:
\begin{equation}\label{jh35kdj}
\frac{dq}{dz}=\frac{j-2q^2-q}{1+z}\,.
\end{equation}
Today, we have $j_0=\frac{dq}{dz}|_{0}+2q_0^2+q_0$. Since we measure that $-1\leq q_0<-1/2$, we clearly obtain: $2q_0^2+q_0>0$ and so, when $q_0 < -1/2$ the term $j_0$ is linked to the sign of the variation of $q$. Although highly predictive, cosmography is jeopardized by some drawbacks which limit its use. In particular:
\begin{description}
  \item[Degeneracy between coefficients]
  The whole list of independent parameters is $q_0,j_0,s_0,\ldots$, but measuring $ H_0$ leads to degeneracy since it is unfortunately 
  impossible to estimate $ H_0$ alone by using measurements of distance modulus. In other words, $H_0$ degenerates with the rest of the parameters.

  \item[Degeneracy with spatial curvature] Spatial curvature  must be fixed somehow, otherwise $j$ and $s$ parameters may be strongly influenced by its value,
  since spatial curvature degenerates with them.
  However, small deviations do not influence the simplest case $\Omega_k=0$ and we extensively use the former case.

  \item[Systematics due to truncated series and convergence]
  Slower convergence in the best fit algorithm may be induced by choosing truncated series at a precise order,
  while systematics in measurements occur, on the contrary, if series are expanded up to a certain order. On the other side, the problem of truncated series lies also on determining the particular Taylor expansion which is better to use with precise data survey. Taylor series, however, are always evaluated around present time\footnote{A complete theory of cosmography at high redshift is today object of debate.}, or better defined when $z=0$. So all data sets exceed the bound $z\simeq 0$ giving, in principle, that all Taylor series do not converge when $z\gg 1$. Combining different data would alleviate the systematics which is produced with the aforementioned theoretical problems, but cannot be considered exhaustive to handle high redshift data sets.

  \end{description}

While degeneracies can be alleviated by refined analyses and combined techniques of cosmographic reconstructions, systematics and convergence are difficult to treat. In particular, to alleviate the convergence problem one can employ parameterizations of the redshift $z$, by means of \emph{auxiliary variables}  ($\mathcal Z_{new}$).
This technique seems to enlarge the convergence radius of Taylor's expansions to a wider sphere having radius $\mathcal Z_{new}<1$ \cite{Cattoen:2007sk}. In such a picture, data lying within $z\in[0,\infty)$ are rewritten into shorter  (non-divergent) ranges. For example, $\mathcal Z_{new} = \frac{z}{1+z}$ whose limits impose $\mathcal Z_{new}\in [0,1]$.
Another technique, more recent and likely suitable, is to consider rational approximations, such as the Pad\'e approximants,
in which the expansions are taken by rational functions which do not diverge as the redshift increases \cite{Gruber:2013wua}.
In order to reduce and alleviate systematics over cosmographic measurements, we show below our new scenarios which refine cosmographic results.
We demonstrate that  using our method the numerics can be clearly improved in a concise and suitable way.

\end{section}

\begin{section}{New strategies toward cosmography with Hubble expansions}  \label{sec:model}

We start by expanding the Hubble function in Taylor series about redshift $z=0$,
\begin{equation} \label{Eofz}
 E(z) \equiv \frac{H(z)}{H_0} = \sum_i \frac{1}{i!}E_i z^i
\end{equation}
with $E_i = H^{(i)}(z)/H_0 |_{z=0}$. The first four coefficients, here named the \emph{eis},
can be written in terms of the statefinders parameters $(q_0,j_0,s_0)$ as
\begin{eqnarray} \label{eisOfsf}
 E_0 &=& 1, \nonumber\\ E_1 &=& 1 + q_0,  \nonumber\\ E_2 &=& -q_0^2 + j_0, \\
   E_3 &=& 3 q_0^2 + 3 q_0^3 -j_0 (4 q_0+3)-s_0. \nonumber
\end{eqnarray}
The comoving distance is defined as the (comoving) distance a photon travels from a source at a redshift $z$ to us, at $z=0$,
\begin{equation}
 \eta(z) =  \int_{0}^{z} \frac{dz'}{H(z')}.
\end{equation}
In our new approach to cosmography we use directly the Taylor expansion of $H(z)$ in the comoving distance expression
and integrate numerically to obtain the luminosity distance, that is

\begin{equation} \label{dlEis}
 \tilde{d}^{(n)}_L(z) \equiv (1+z)\int_{0}^{z}  \left( \sum_{i=0}^n \frac{1}{i!}E_i z^i \right)^{-1} dz'.
\end{equation}

From Eq.~(\ref{Eofz}) we can note that at low redshifts not all of the powers in the $E(z)$ expansion are important.
Thus, in order to speed up the numerical computations,
we estimate the \emph{eis} parameters in a hierarchical manner. To this end we use Eq.~(\ref{dlEis}) in redshift bins as

\begin{equation} \label{EisModel}
   \tilde{d}_L(z;E_1,E_2,E_3) = \begin{cases}
               \tilde{d}^{(1)}_L(z)               & z< z_{low}\\
               \tilde{d}^{(2)}_L(z)               & z_{low}< z < z_{mid}\\
               \tilde{d}^{(3)}_L(z)               & z_{mid}< z < z_{high}.
           \end{cases}
\end{equation}
For $z>z_{high}$ we expand in Taylor series the integrand of $\tilde{d}^{(3)}_L(z_k)$ up to $z^3$ and analytically perform the integration.
This last step is necessary for numerical stability, otherwise the tails of the posterior distributions become very noisy.

That is, for a supernova at redshift $z_k$ in a given simulated catalog, we use $\tilde{d}^{(1)}_L(z_k)$ if $z_k<z_{low}$,
$\tilde{d}^{(2)}_L(z_k)$ if $z_{low}< z_k < z_{mid}$, and $\tilde{d}^{(3)}_L(z_k)$ if $z_{mid}< z_k < z_{high}$.
We performed preliminar numerics and found that a good choice for these redshift cuts is
\begin{equation}\label{redshiftscuts}
z_{low} = 0.05, \qquad z_{mid} = 0.4, \qquad z_{high} = 0.9.
\end{equation}
The numerical outcomes of this particular binning does not differ significatively from those obtained by a direct application of
Eq.~(\ref{dlEis}).
We name the method of Eq.~(\ref{EisModel}), the \Eis method, or simply \emph{Eis}.

We further propose two hybrid methods:
\emph{hybrid\_1} which consists in the use of SC up to $z_{mid}$, and beyond that redshift in using the \emph{Eis} method with the integrand expanded.
The other method is \emph{hybrid\_2}, which
is a modification of \emph{hybrid\_1} with $z_{high}=z_{mid} = 0.4$.
The physical reasons behind the two hybrid approaches are essentially based on the fact that at low redshifts,
SC constrains with small dispersions the deceleration and jerk parameters.
The here involved equations for the hybrid methods are given in Appendix \ref{app::formulas}.

In the following sections we use the code CosmoMC \cite{Lewis:2002ah} to draw the likelihood distributions for all the methods studied here with a
Metropolis-Hasting MCMC algorithm \cite{Hastings:1970aa}. A module for CosmoMC is available at
\href{https://github.com/alejandroaviles/EisCosmography}{https://github.com/alejandroaviles/EisCosmography}; also, all the
simulated catalogs and further statistics can be found there.

Despite our numerics perform fits of the \eis parameters (even for SC; see Eq.~(\ref{dlSCeis})), we are primarily interested in the statefinders
because its physical interpretation is more familiar. We are imposing flat priors on the \emph{eis} parameters, which is not equivalent to flat priors on the
statefinders. Nevertheless, this does not have a significative impact, as can be observed by comparing
the triangular plot in Fig.~\ref{fig:1} (bottom panel) to
previous works on cosmography; see, {\it e.g.} \cite{Aviles:2012ay,Busti:2015xqa}.

\end{section}

\begin{section}{Numerical analysis with simulated data}  \label{sec:sdanalysis}

In this section we test the performance of our \emph{Eis} method by using simulated catalogs of supernovae Ia. To this end, we construct
two kinds of simulated data based on fiducial $\Lambda$CDM and wCDM models.
On each of these simulations we take
740 data distributed with the same redshifts and error bars of the observed peak magnitudes ($\sim 0.12$)
as those of the JLA compilation \cite{Betoule:2014frx}.
We choose this catalog because it has a large amount of
low redshift data providing a good inference of $E_1$ that acts as a leverage for a better estimation of the rest of \eis parameters. We  better discuss 
this point in Sect.~(\ref{Subsec:RSdist}).

\begin{subsection}{The dispersed simulated data}  \label{subsec:sdanalysis}

\begin{figure}
\begin{center}
\includegraphics[width=3in]{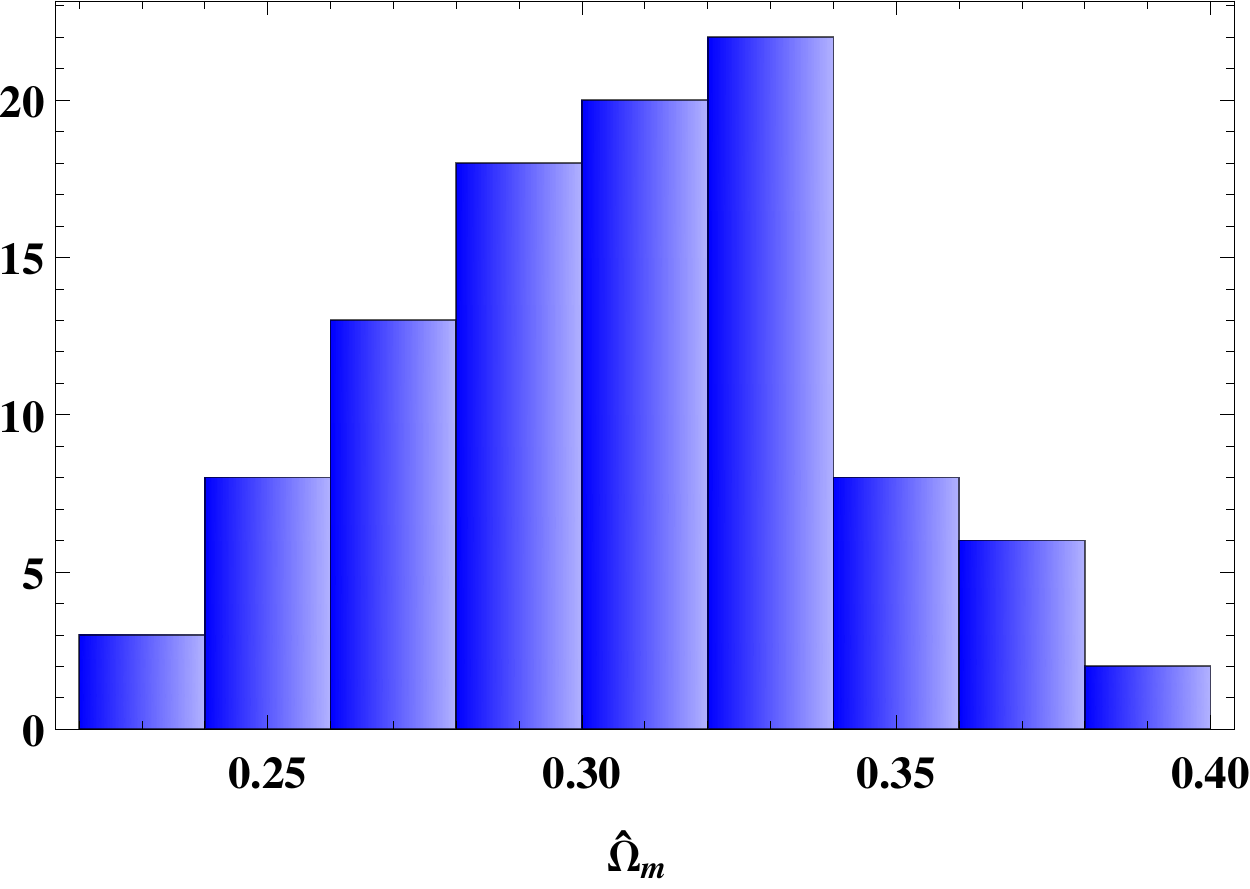}
\caption{Histogram of the estimations $\hat{\Omega}_m$ for the 100 supernovae simulated catalogs.}
\label{histOm}
\end{center}
\end{figure}

\begin{figure}
\begin{center}
\includegraphics[width=3.2in]{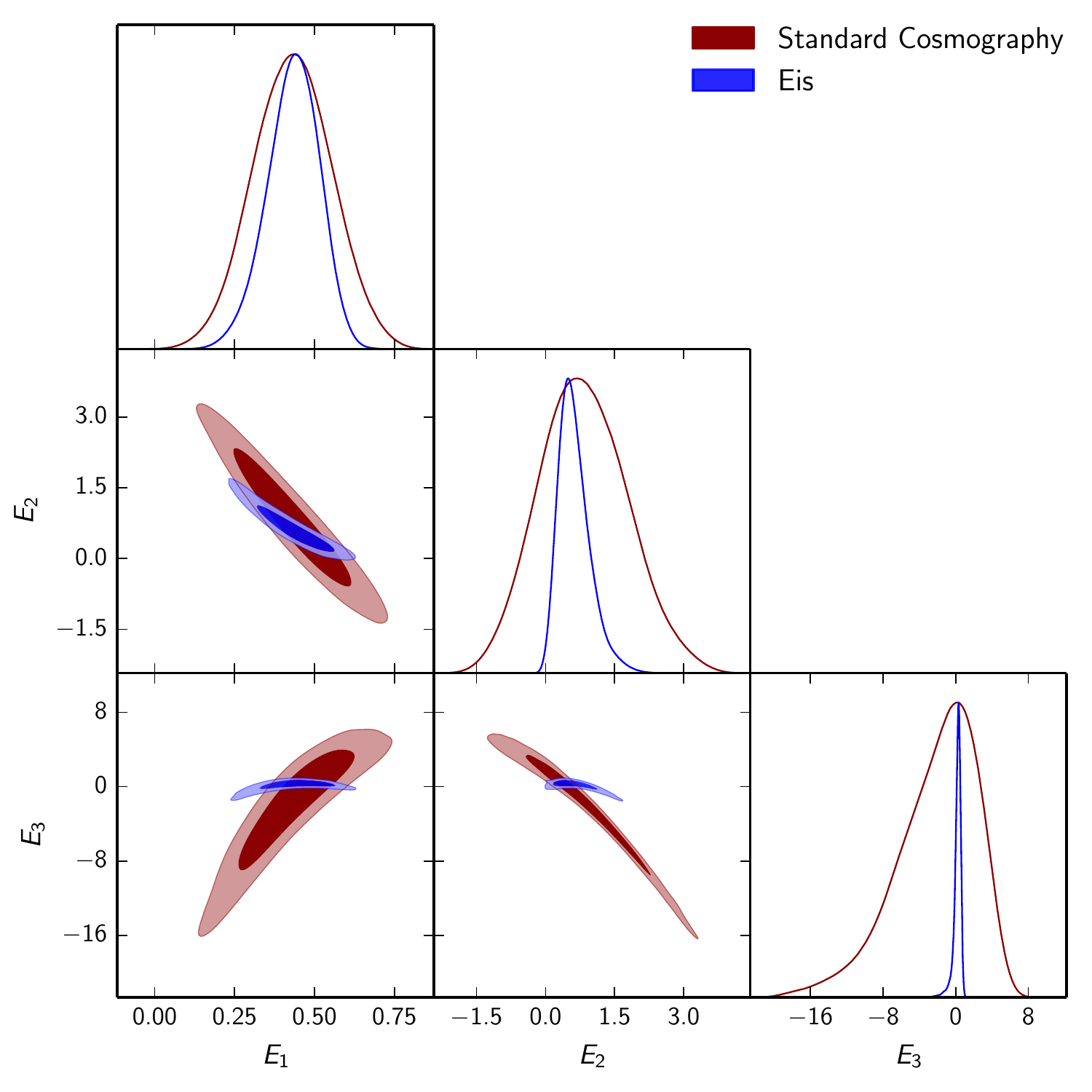}
\includegraphics[width=3.2in]{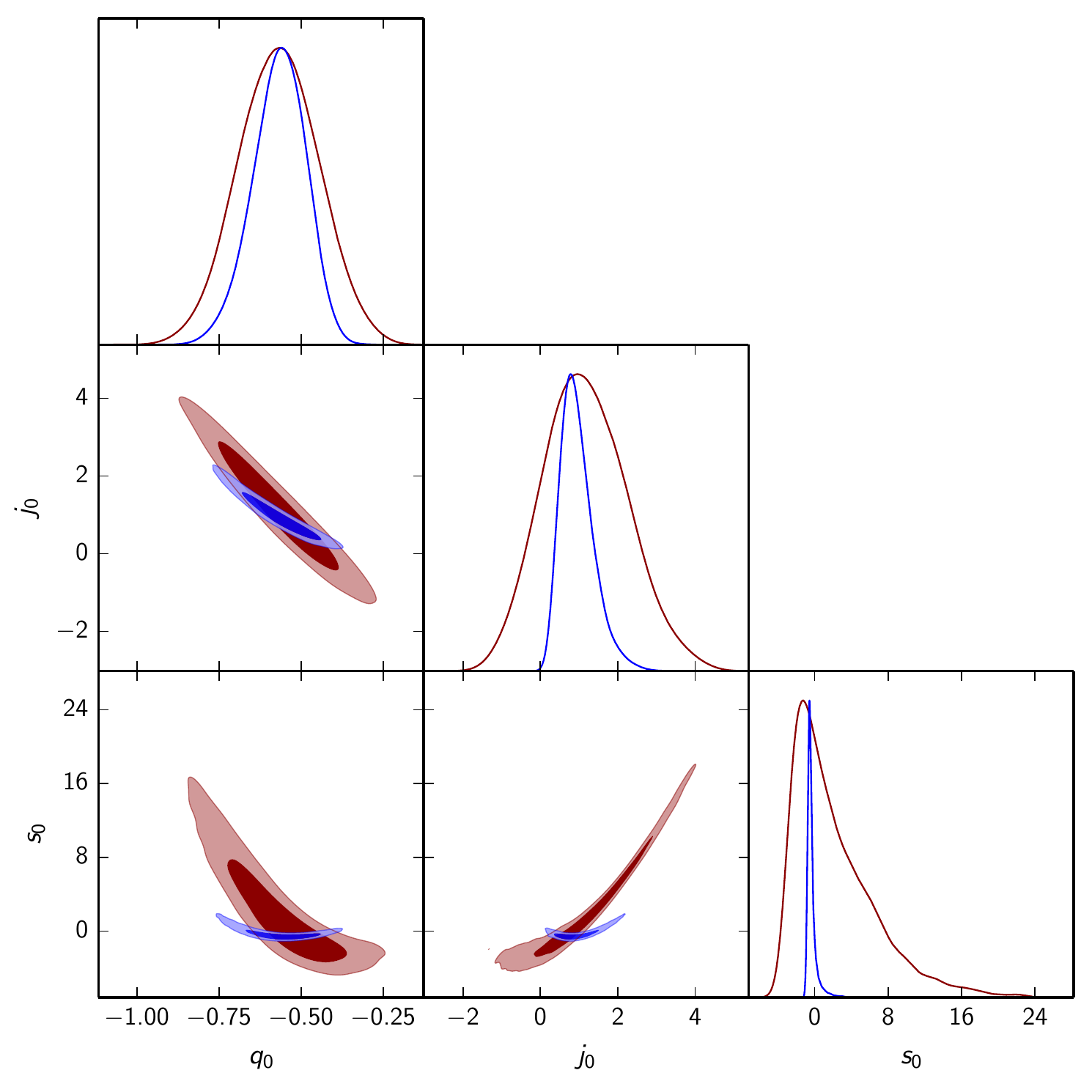}
\caption{Triangle plots for the estimated \emph{eis} parameters  (top) and the derived statefinder parameters (bottom).
This is for the simulated catalog \emph{SD\_1}.}
\label{fig:1}
\end{center}
\end{figure}

\begin{figure}
\begin{center}
\includegraphics[width=3in]{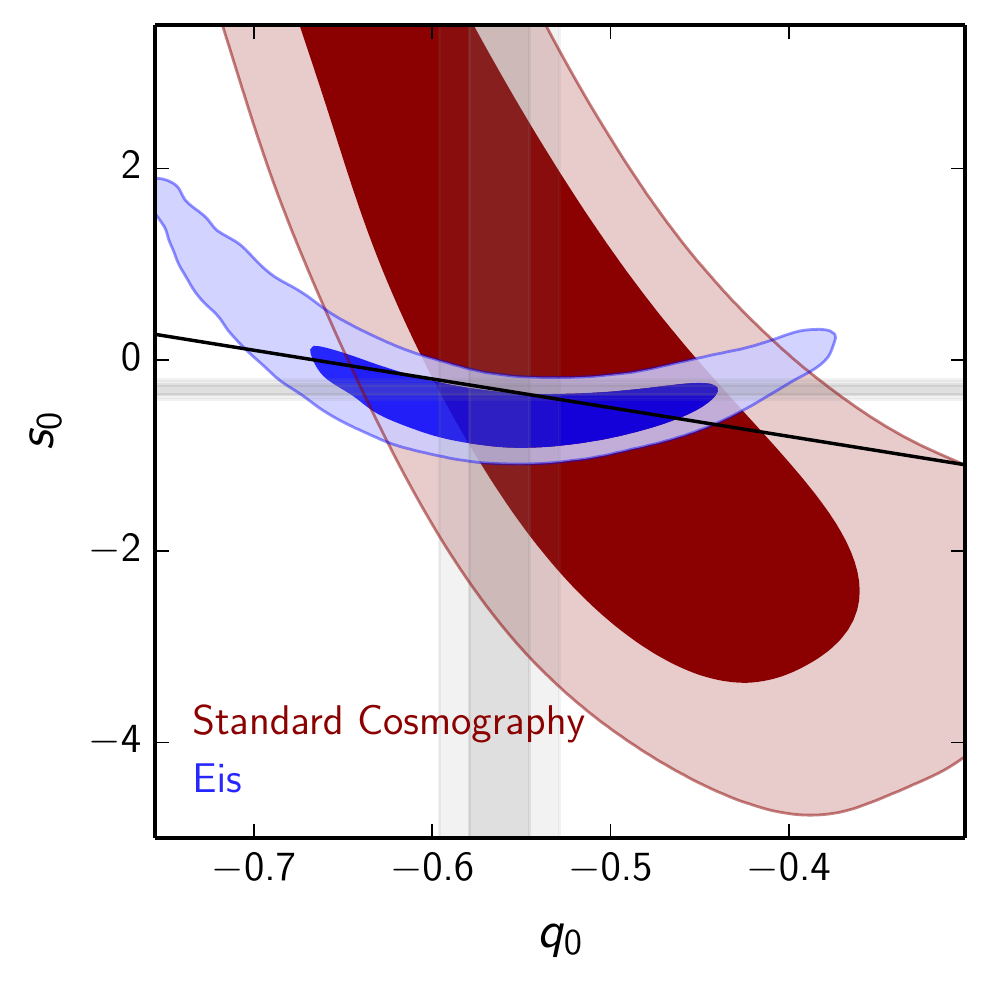}
\caption{Zoom of the $q_0$-$s_0$ contour plot of Fig.~\ref{fig:1}. We also show the region allowed by the flat $\Lambda$CDM model (solid line) and the
$\Lambda$CDM confidence intervals (horizontal and vertical shadows).}
\label{fig:1_1}
\end{center}
\end{figure}

\begin{figure}
\begin{center}
\includegraphics[width=3in]{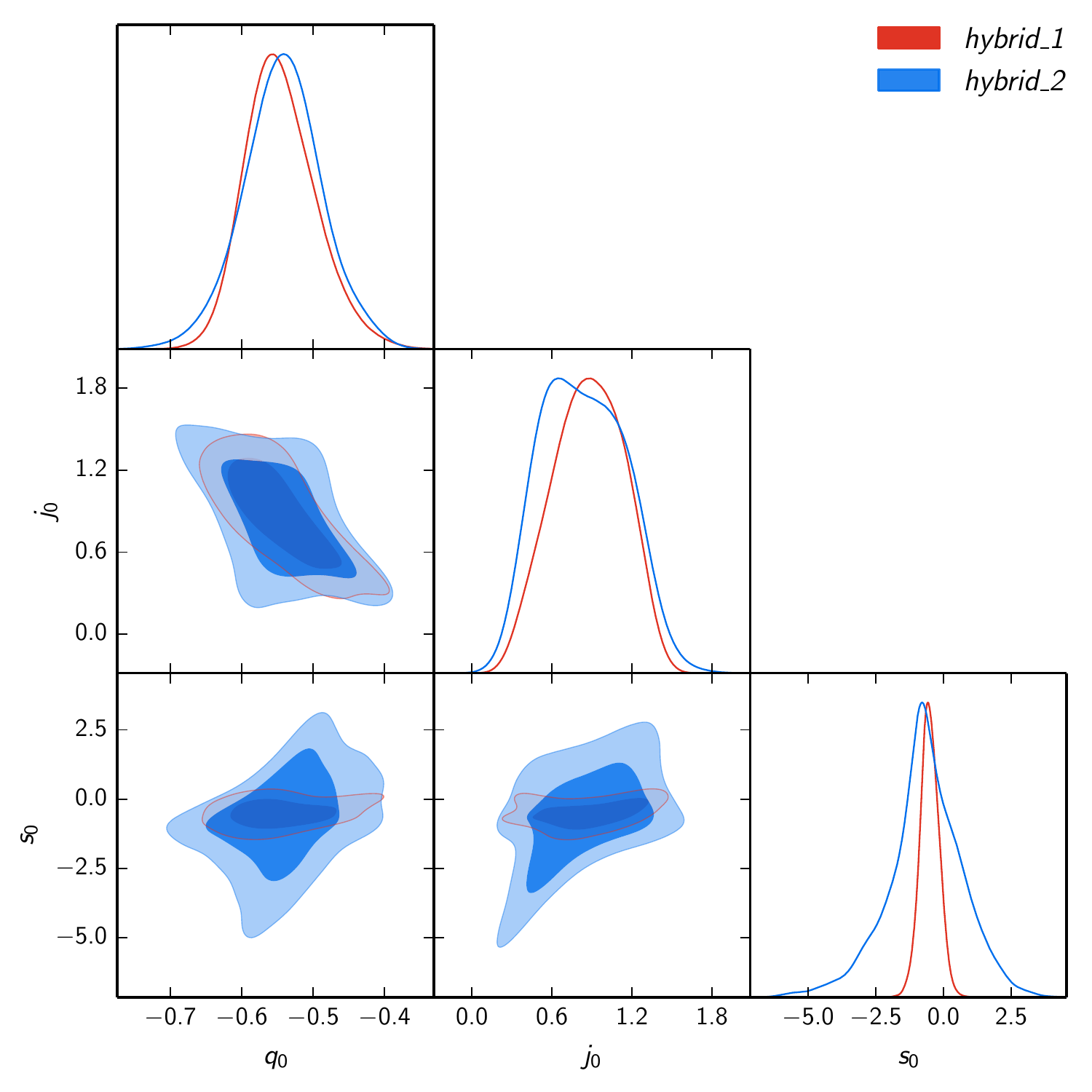}
\caption{Triangular plots for \emph{hybrid\_1} (red) and \emph{hybrid\_2} (blue) methods, obtained by fitting to the \emph{SD\_1} simulated
catalog.}
\label{fig:tri_hybrids}
\end{center}
\end{figure}

\begingroup
\squeezetable
\begin{table*}
\caption{Marginalized 1D estimations using simulated data for the \emph{SD\_1} catalog. See text for details.
The complete table for the 100 catalogs, as well as one for the \emph{eis} parameters, can be found in
\href{https://github.com/alejandroaviles/EisCosmography}{https://github.com/alejandroaviles/EisCosmography}.}
\begin{tabular}{@{}llrrrcrrrcrrrrccccc@{}}
\toprule
             &
\phantom{ab} & \multicolumn{3}{c}{$q_0$} &
\phantom{ab} & \multicolumn{3}{c}{$j_0$}&
\phantom{ab} & \multicolumn{3}{c}{$s_0$}&
\phantom{ab}  & Bias  &&&\\
\cmidrule{3-5} \cmidrule{7-9} \cmidrule{11-13}&
             &  {\tiny Mean ($\sigma$)} & \tiny{ \emph{0.68 c.l.}} & \tiny{ \emph{0.95 c.l.}} &&
                {\tiny Mean ($\sigma$)} & \tiny{ \emph{0.68 c.l.}} & \tiny{ \emph{0.95 c.l.}} &&
                {\tiny Mean ($\sigma$)} & \tiny{ \emph{0.68 c.l.}} & \tiny{ \emph{0.95 c.l.}} &
\phantom{ab}  & $\Delta \chi^2 $ & \phantom{ab} & FoM &\\

\midrule \\[-3pt]

\emph{Eis}&&$-0.567$ ($ 0.082)$ & ${}^{+0.089}_{-0.073}$ & ${}^{+0.150}_{-0.173}$ &&
               $ 0.960$ ($ 0.446)$ & ${}^{+0.289}_{-0.517}$ & ${}^{+0.906}_{-0.773}$ &&
               $-0.318$ ($ 0.597)$ & ${}^{+0.091}_{-0.498}$ & ${}^{+1.134}_{-0.727}$ &&    0.923&& 0.0107&\\[5pt]

SC          && $-0.568$ ($ 0.122)$ & ${}^{+0.124}_{-0.124}$ & ${}^{+0.240}_{-0.241}$ &&
               $ 1.161$ ($ 1.095)$ & ${}^{+1.025}_{-1.169}$ & ${}^{+2.155}_{-2.086}$ &&
               $ 2.147$ ($ 4.565)$ & ${}^{+1.820}_{-5.296}$ & ${}^{+9.443}_{-6.395}$ &&    0.920&& 0.1201&\\[5pt]

$\Lambda$CDM &&$-0.562$ ($ 0.017)$ & ${}^{+0.017}_{-0.017}$ & ${}^{+0.033}_{-0.033}$ &&
               $\,\,1$ &&&&
               $-0.313$ ($ 0.051)$ & ${}^{+0.050}_{-0.050}$ & ${}^{+0.098}_{-0.098}$ && -- && --& \\[5pt]

\emph{hybrid\_1}&&$-0.545$ ($ 0.051)$ & ${}^{+0.042}_{-0.056}$ & ${}^{+0.104}_{-0.094}$ &&
               $ 0.875$ ($ 0.269)$ & ${}^{+0.303}_{-0.258}$ & ${}^{+0.481}_{-0.527}$ &&
               $-0.517$ ($ 0.358)$ & ${}^{+0.340}_{-0.350}$ & ${}^{+0.704}_{-0.709}$ &&    0.908&& 0.0085&\\[5pt]

\emph{hybrid\_2}&&$-0.543$ ($ 0.057)$ & ${}^{+0.054}_{-0.054}$ & ${}^{+0.115}_{-0.115}$ &&
               $ 0.835$ ($ 0.315)$ & ${}^{+0.307}_{-0.361}$ & ${}^{+0.581}_{-0.560}$ &&
               $-0.684$ ($ 1.432)$ & ${}^{+1.465}_{-1.127}$ & ${}^{+2.746}_{-3.194}$ &&     0.179&& 0.0367&\\[5pt]

\bottomrule
\end{tabular}
\label{Table:longtable}
\end{table*}
\endgroup

The dispersed simulated data set consists in 100 simulations in which each supernovae in every catalog is obtained from fiducial $\Lambda$CDM models
with $\Omega_m$ drawn from a Gaussian distribution with mean $0.30$ and standard deviation
of $0.034$.\footnote{We fix the Hubble constant to $H_0 = 70 \, \rm{km}/\rm{s}/\rm{Mpc}$. Nevertheless, it is irrelevant because supernovae data
cannot estimate the Hubble constant. Therefore, in the numerical analysis we
internally marginalize the combination $5\ln(c/H_0) + M_b$ as described in \cite{Goliath:2001af}.}
That is, in a single catalog, for each supernova at redshift $z_k$ we give the modulus distance a value
$\mu_k = \mu(z_k;\Omega_{m\,k} \in \mathcal{N}(0.30,0.034))$. Taking the mean value and the standard deviation of this $\Omega_m$ distribution and using
$\Lambda$CDM-statefinder relations, we expect that our results intersect the intervals
$q_0 = -0.55 \pm 0.051$, $j_0 = 1$ and $s_0 = -0.35 \pm 0.153$. This is exactly the case that we got for all the models involved into our analyses.

One may be tempted to assume that the underlying, {\it true}, cosmology of the dispersed simulated data sets is the same for all of them, and given
by $\Omega_m = 0.3$; see, {\it e.g.} \cite{Busti:2015xqa}.
The drawback of this approach is that the true cosmology is actually unknown for each catalog.
Thus, in order to analyze how good the fits are, we must compare against $\Lambda$CDM fittings to the same simulations.

In Fig.~\ref{histOm} we show a histogram of the estimated $\hat{\Omega}_m$'s by performing these fits. The average value is
$\langle \hat{\Omega}_m \rangle = 0.307$, the dispersion is $\sigma_{\hat{\Omega}_m} = 0.036$, and the average of the standard deviations is
$\langle \sigma_{\Omega_m} \rangle = 0.012$.

In Fig.~\ref{fig:1} we show a triangular plot for the derived statefinders of one of our simulated catalogs, named \emph{SD\_1},
both for the SC and
the \emph{Eis} methods, revealing that the dispersions are smaller than in SC, most notably for the parameter $E_3$ (or $s_0$). For the $E_1$
parameter there is also an improvement over SC, nevertheless this is not as remarkable as for the other parameters,
the reason behind this is that $q_0$ is constrained mainly by low redshift supernovae, as it is shown in Sec.\ref{Subsec:RSdist} below,
particularly in Fig.~\ref{fig:fixtolcdmZBins}, and all the expansions considered in this work converge at low redshift.
In Fig.~\ref{fig:1_1} we display a zoom of the $q_0$-$s_0$ 2D joint posterior. The shadows show the confidence intervals (\emph{c.i.})
for $q_0$ and $s_0$ derived from fitting the
$\Lambda$CDM model. The solid black line is $s_0 = -2-3 q_0$, which corresponds to the allowed region
in $\Lambda$CDM [this can be derived by setting $w=-1$ in Eqs.~(\ref{app:sflcdm})]. We notice that SC does not follow this degeneracy trend, while
\emph{Eis} can do it inside its $0.68$ confidence region.

Complementing Fig.~\ref{fig:1}, in Table \ref{Table:longtable} we show the 1-dimensional marginalized posterior intervals for the simulated data \emph{SD\_1}.
The average statistics of the standard deviations and mean posterior
values for all the methods are shown in Table \ref{Table::AverageStats}.

In Fig.~\ref{fig:tri_hybrids} we show the triangular plot derived from the fitting to the simulated catalog \emph{SD\_1}. We make note
that the \emph{hybrid\_1} method provides the smallest dispersions among the four methods studied here.

From the results of this subsection, we conclude that the approaches proposed in this work improve the standard deviations of the
statefinders estimations from those obtained by SC.

\begingroup
\squeezetable
\begin{table*}
 \caption{Average statistics for the 100 simulated catalogs.}

\begin{tabular}{@{}lcrrrrc@{}}
\toprule
& \\[-2pt]
& \phantom{abcdfghijklmno}        & \phantom{abcab} \emph{Eis}
                              & \phantom{abcab} \emph{hybrid\_1} & \phantom{abcab} \emph{hybrid\_2} & \phantom{abcab} \phantom{abcd} \emph{SC} &\phantom{ab}\\[3pt]
\midrule
& \\[-2pt]

&$\langle \hat{E}_1 \rangle$	        & $ 	0.458	$ & $	0.474	$ & $	0.476	$ & $	0.457	$& \\[2pt]
&  $\langle \sigma_{E_1} \rangle$       & $ 	0.082	$ & $	0.052	$ & $	0.056	$ & $	0.124	$& \\[2pt]
&  $\sigma_{\hat{E}_1}$                 & $ 	0.060   $ & $	0.050	$ & $	0.046	$ & $	0.057	$& \\[2pt]
&  $\langle b_{E_1} \rangle$            & $ 	-0.003	$ & $	0.013	$ & $	0.016	$ & $	-0.004	$& \\[2pt]
&  $\langle risk(E_1)\rangle$           & $ 	0.083	$ & $	0.054	$ & $	0.059	$ & $	0.125	$& \\[6pt]

&$\langle \hat{E}_2 \rangle$	        & $ 	0.634	$ & $	0.596	$ & $	0.578	$ & $	0.830	$& \\[2pt]
&  $\langle \sigma_{E_2} \rangle$       & $ 	0.351	$ & $	0.247	$ & $	0.288	$ & $	0.964	$& \\[2pt]
&  $\sigma_{\hat{E}_2}$                 & $ 	0.014	$ & $	0.065	$ & $	0.110	$ & $	0.033	$& \\[2pt]
&  $\langle b_{E_2} \rangle$            & $ 	-0.073	$ & $	-0.110	$ & $	-0.129	$ & $	0.124	$& \\[2pt]
&  $\langle risk(E_2) \rangle$          & $ 	0.124	$ & $	0.121	$ & $	0.142	$ & $	0.181	$& \\[6pt]

&$\langle \hat{E}_3 \rangle$            & $ 	0.191	$ & $	0.282	$ & $	0.368	$ & $	-2.024	$& \\[2pt]
&  $\langle \sigma_{E_3} \rangle$       & $ 	0.450	$ & $	0.452	$ & $	1.579	$ & $	4.479	$& \\[2pt]
&  $\sigma_{\hat{E}_3}$                 & $ 	0.141	$ & $	0.037	$ & $	0.222	$ & $	0.302	$& \\[2pt]
&  $\langle b_{E_3} \rangle$            & $ 	0.256	$ & $	0.346	$ & $	0.432	$ & $	-1.960	$& \\[2pt]
&  $\langle risk(E_3) \rangle$          & $ 	0.311	$ & $	0.351	$ & $	0.436	$ & $	1.964	$& \\[6pt]

\midrule
& \\[-2pt]
&  $\langle \hat{q}_0 \rangle$          & $-0.542$ & $-0.526$ & $-0.524$ & $-0.543$ &\\[2pt]
&  $\langle \sigma_{q_0} \rangle$       & $ 0.082$ & $ 0.052$ & $ 0.056$ & $ 0.124$ &\\[2pt]
&  $\sigma_{\hat{q}_0}$                 & $ 0.060$ & $ 0.050$ & $ 0.046$ & $ 0.057$ &\\[2pt]
&  $\langle b_{q_0} \rangle$            & $-0.003$ & $ 0.013$ & $ 0.016$ & $-0.004$ &\\[2pt]
&  $\langle risk(q_0) \rangle$          & $ 0.082$ & $ 0.054$ & $ 0.058$ & $ 0.125$ &\\[6pt]

&  $\langle \hat{j}_0 \rangle$          & $ 0.938$ & $ 0.878$ & $ 0.857$ & $ 1.144$ &\\[2pt]
&  $\langle \sigma_{j_0} \rangle$       & $ 0.439$ & $ 0.285$ & $ 0.317$ & $ 1.095$ &\\[2pt]
&  $\sigma_{\hat{j}_0}$                 & $ 0.074$ & $ 0.017$ & $ 0.061$ & $ 0.063$ &\\[2pt]
&  $\langle b_{j_0} \rangle$            & $-0.062$ & $-0.122$ & $-0.143$ & $ 0.144$ &\\[2pt]
&  $\langle risk(j_0) \rangle$          & $ 0.450$ & $ 0.310$ & $ 0.352$ & $ 1.106$ &\\[6pt]

&  $\langle \hat{s}_0 \rangle$          &  $-0.432$ & $-0.642$ & $-0.731$ & $ 1.985$\\[2pt]
&  $\langle \sigma_{s_0} \rangle$       &  $ 0.590$ & $ 0.351$ & $ 1.366$ & $ 4.424$\\[2pt]
&  $\sigma_{\hat{s}_0}$                 &  $ 0.361$ & $ 0.240$ & $ 0.069$ & $ 0.545$\\[2pt]
&  $\langle b_{s_0} \rangle$            &  $-0.050$ & $-0.260$ & $-0.349$ & $ 2.367$\\[2pt]
&  $\langle risk(s_0) \rangle$          &  $ 0.628$ & $ 0.445$ & $ 1.414$ & $ 5.018$\\[5pt]

\bottomrule
\end{tabular}
\label{Table::AverageStats}
\end{table*}
\endgroup

\end{subsection}

\begin{subsection}{Bias on the estimators}

We use the bias of an estimator $\hat{\theta}$  defined as
\begin{equation}
 b_{\theta} = \text{bias}(\hat{\theta}) \equiv \hat{\theta} - \theta_{\text{true}}
\end{equation}
where $\theta_{\text{true}}$ is the true value of the parameter $\theta$. For the estimated $\hat{\theta}$ we use the mean value
of the posterior distribution. We may use the maximum likelihood estimator
instead, but our preliminary numerics show only slight differences despite the distributions are in general skewed.

As explained above, we do not know the true values of the parameters $\theta$; thus, we will assume that
the $\Lambda$CDM provides unbiased estimations for them. That is, we use $\theta_{\text{true}}=\hat{\theta}_{\Lambda\text{CDM}}$.

By itself, the bias does not provide a complete information of how well $\hat{\theta}$ estimates
$\theta$, specially if we know only approximately the true value. For this reason, it is convenient to use complementary statistics.
First, we implement the risk statistics \cite{kendalls,Linder:2008qj}
\begin{equation} \label{rs1}
 risk(\theta) = \sqrt{\sigma_\theta^2 + b_\theta^2},
\end{equation}
which penalize the bias with the standard deviation. Furthermore, for the whole cosmology, following \cite{kendalls,Press:1992:NRC:148286}
(see also \cite{Samsing:2009kb,Shapiro:2008yk} for applications in cosmology), we compute the bias statistics
\begin{equation} \label{bsDchi2}
 \Delta \chi^2 =  \mathbf{b}^{\text{T}} \mathbf{F} \,\mathbf{b},
\end{equation}
which roughly quantifies the slip from the $\chi^2$-statistics due to bias.
Here $\mathbf{F}$ is the reduced Fisher matrix for the estimated parameters and $ \mathbf{b} = (b_{E_1},b_{E_2},b_{E_3}) $ is the bias
vector for \emph{eis} parameters.\footnote{$\Delta \chi^2$ is not an invariant between statefinders ($sf$) and \emph{eis} parameters because they
are not linearly related and therefore $\widehat{sf}\ne sf(\widehat{eis})$. Nevertheless, the differences in values are small. On the other hand,
the determinant of the Fisher matrix, and hence the FoM, is an invariant because the
determinant of the transformation matrix between $eis$ and $sf$ parameters is equal to $-1$.}
We note that this statistics requires the maximum likelihood estimator instead of the posterior
mean.

A smaller $\Delta \chi^2$ does not imply a smaller bias, this can be noted for the case of one single parameter,
since $\Delta \chi^2 = b_\theta^2/\sigma_\theta^2$ for this case.
Therefore,  we additionally compute the figure of merit (FoM), that we define as
\begin{equation}
 \text{FoM} = \frac{4 \pi}{3} \frac{1}{\sqrt{\det \mathbf{F}}}.
\end{equation}
The numerical factor $4 \pi / 3$ is not common in the literature. We consider it so that the FoM coincides with the volume of
the 3-dimensional ellipsoid defined by the covariance matrix.

Strictly, the FoM and $\Delta \chi^2$ bias statistics work only for multi-variate Gaussian distributions. By using them, we are
implicitly approximating the confidence regions by ellipsoids.
In Fig.~\ref{fig:ellipse} we show the projection for two of these on the $q_0\text{-}s_0$ subspace. For comparison we also show
the Markov chains obtained in the MCMC analysis, as well as the region allowed in the $\Lambda$CDM model.

We perform the four statistics for each simulated catalog and for each one of our models.

In Table \ref{Table::AverageStats} we show the average values of the bias and risk for both the \eis and statefinder parameters.
It can be noted that in these 1-parameter bias tests, \emph{Eis}, as well as \emph{hybrid\_1} and \emph{hybrid\_2}, perform better than SC.

For the whole 3-dimensional bias statistics, the average values over the 100 simulated catalogs for
$\Delta \chi^2$ and FoM  are:

\begin{align} \label{dchi2_averages}
&\text{SC:}                &\langle \Delta \chi^2 \rangle =0.806&,         &  \text{FoM} = 0.1226.         \nonumber\\
&\text{\emph{Eis}:}        &\langle \Delta \chi^2 \rangle =2.121&,         &  \text{FoM} = 0.0108.         \nonumber\\
&\text{\emph{hybrid\_1}:}  &\langle \Delta \chi^2 \rangle =1.989&,         &  \text{FoM} = 0.0095.         \nonumber\\
&\text{\emph{hybrid\_2}:}  &\langle \Delta \chi^2 \rangle =0.168&,         &  \text{FoM} = 0.0378.
\end{align}
For $\nu=3$ parameters, the $1\sigma$ contour is $\Delta \chi^2 \leq 3.53$ \cite{Press:1992:NRC:148286}, thus for all the methods,
on the average, the true value is well inside this region. We note SC is able to do it because the volume of its error ellipsoid, or FoM, is
very large in comparison with the other methods; cf. Fig.~\ref{fig:ellipse}.
We have seen in Fig.~\ref{fig:01} that, when using 2 parameters, the true
value lies outside the SC $2\sigma$ confidence region. This is because when the number of parameters is reduced, the bias is propagated from higher to lower order
cosmographic parameters. In the upcoming section we will observe this effect more clearly by reducing the number of parameters to just one.

\begin{figure}
\begin{center}
\includegraphics[width=2.5in]{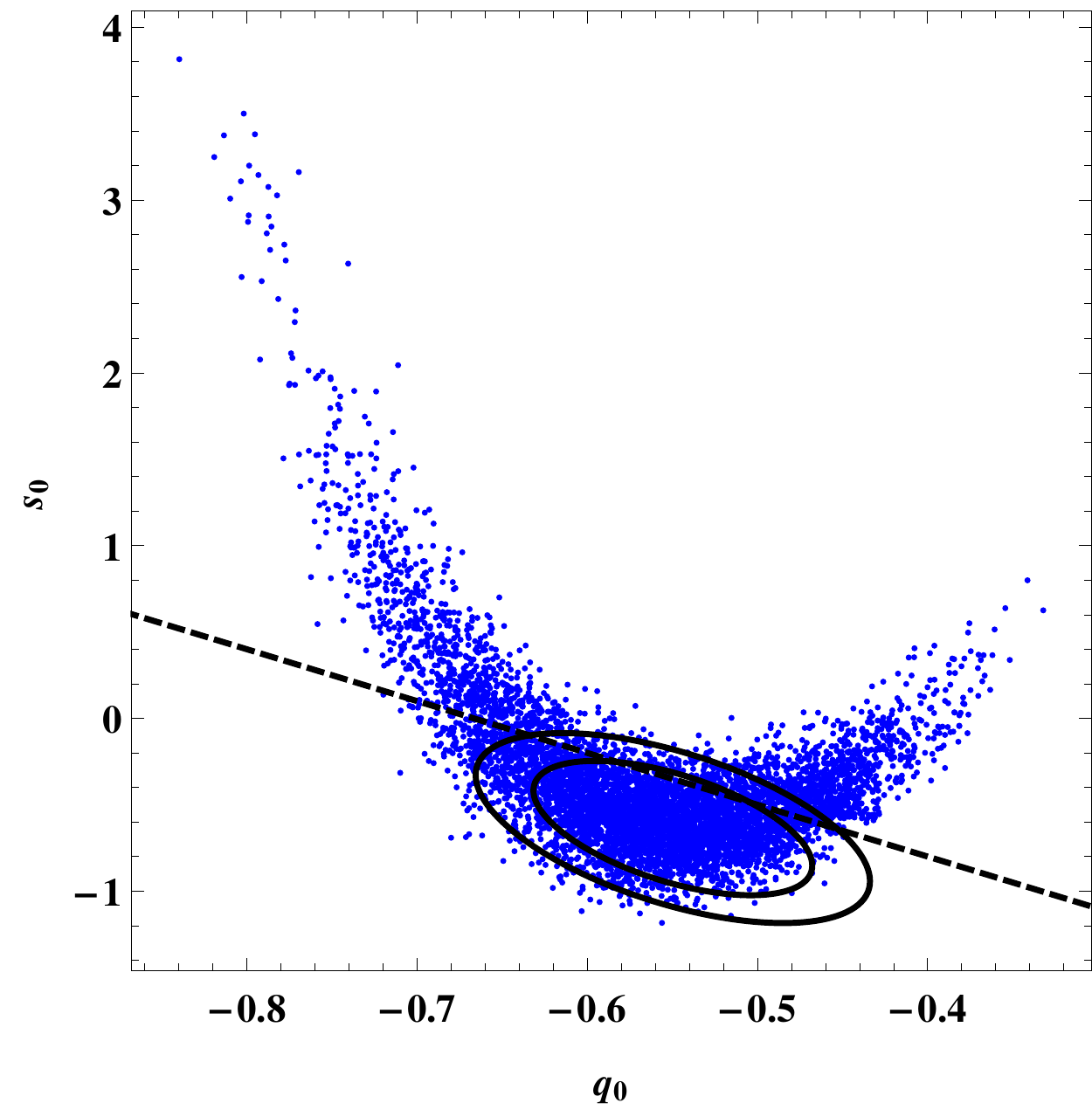}
\includegraphics[width=2.5in]{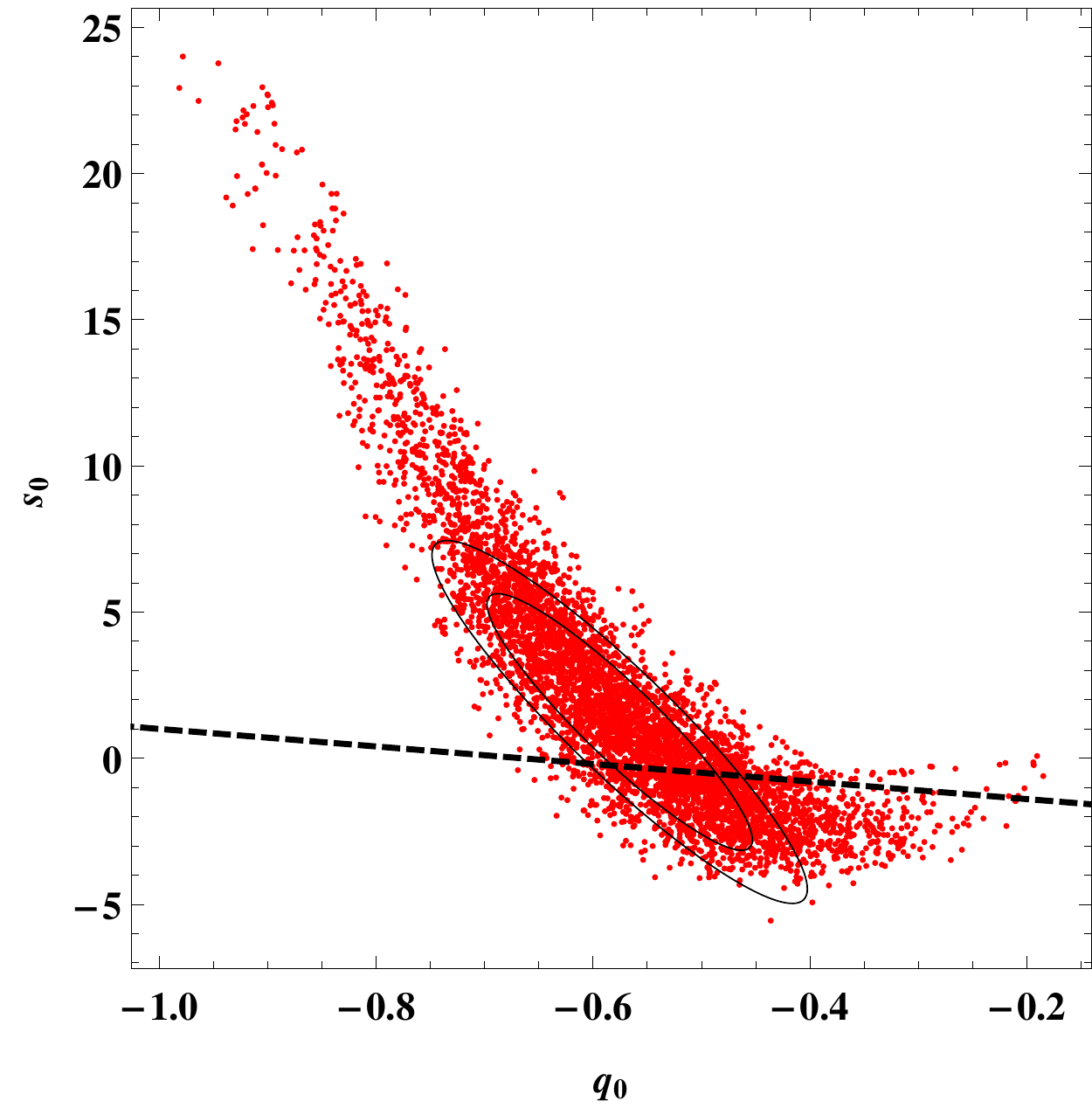}
\caption{Multivariate Gaussian approximation used by the figure of merit and bias statistic $\Delta \chi^2$ tests. The up (bottom) panel
is for the \emph{Eis} (SC) method. The solid lines
show the ellipses at $1$ and  $2 \sigma$. The points are the drawn Markov Chain in the MCMC \emph{SD\_1}. For reference we show
degeneracy line in $\Lambda$CDM (dashed lines). Note the scales on the $s_0$ axes.}
\label{fig:ellipse}
\end{center}
\end{figure}

On the overall analysis it seems the three new methods presented in this work have a similar performance.
Thus, by virtue of the simplicity of \emph{Eis} and the
more appealing form of its contour plots, in
the following we focus mainly to this method.

\end{subsection}

\begin{subsection}{The exact simulated data} \label{subsec:exactSD}

\begin{figure}
\begin{center}
\includegraphics[width=2.5in]{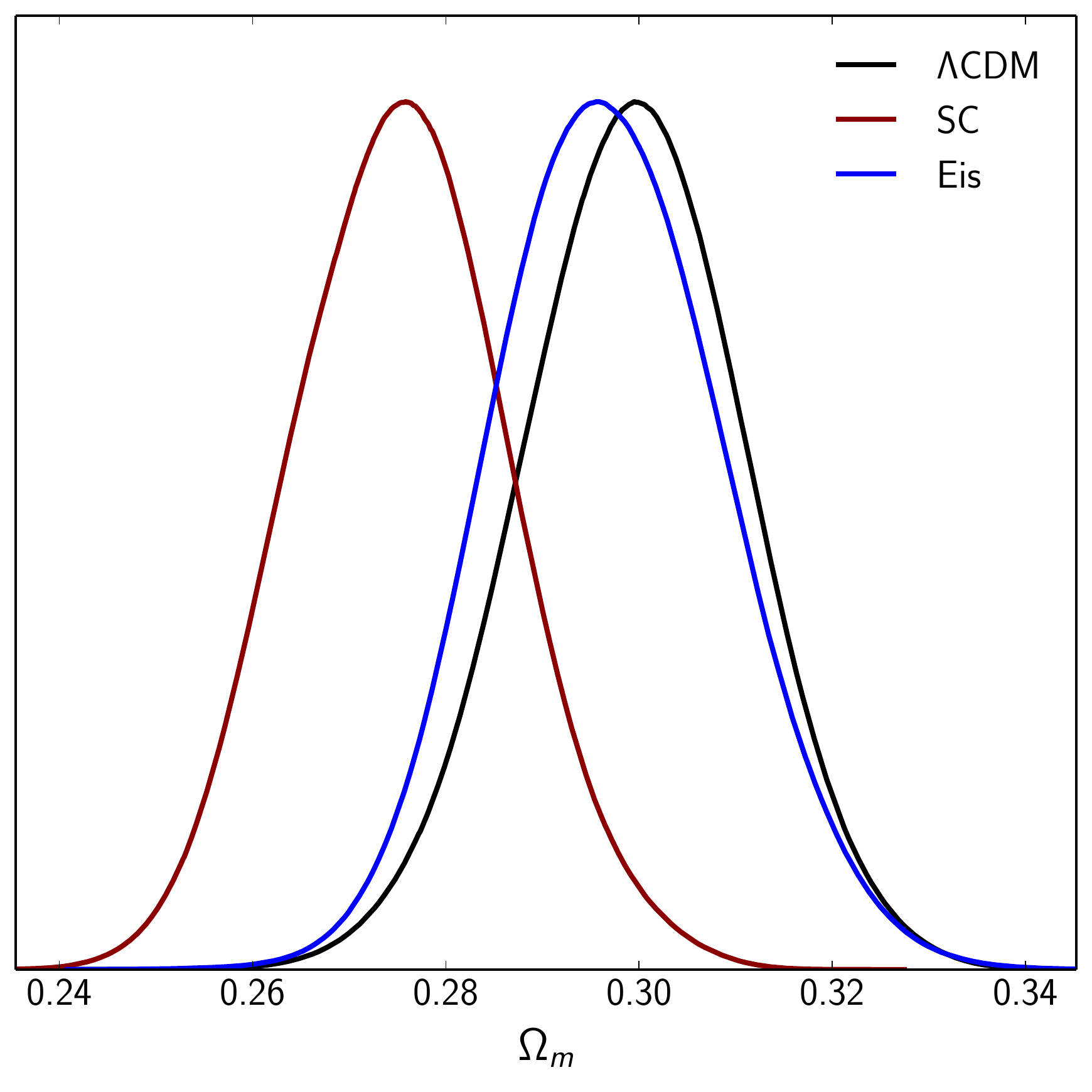}
\includegraphics[width=2.5in]{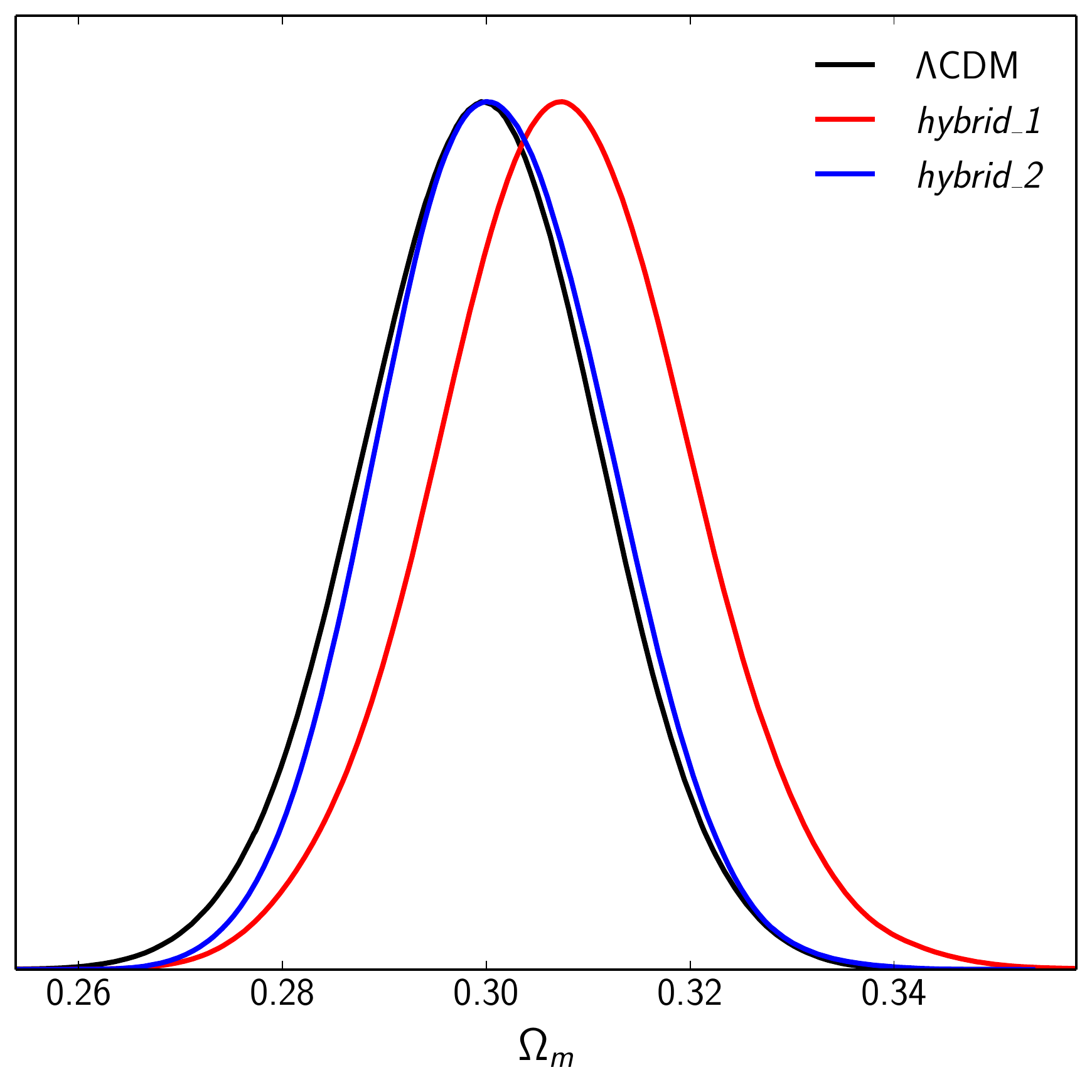}
\caption{One-parameter diagnostic. We show the estimated $\Omega_m$ for $\Lambda$CDM (black curve), SC (red curve) and \emph{Eis} (blue curve)
methods (top figure); and for $\Lambda$CDM (black curve), \emph{hybrid\_1} (red curve) and \emph{hybrid\_2} (blue curve)
methods (bottom figure). We let
vary $E_1$  and fix the rest \emph{eis} parameters with the degeneracies in $\Lambda$CDM. See text for details.}
\label{fig:fixtolcdm}
\end{center}
\end{figure}

The set of \emph{exact} simulated catalogs consists in 25 simulations based on $w$CDM models with
$\Omega_m=(0.25, 0.28, 0.30, 0.32, 0.35)$ and equation of state parameters for dark energy  $w=(-1.10, -1.05, -1, -0.95, 0.90)$.
On each of these simulations we take
740 data distributed with the same redshifts and error bars as the JLA compilation, and for each supernovae at redshift $z_k$ we attribute the exact value
$\mu_k=\mu(z_k;\Omega_{m}, w)$ to the modulus distance.  The aim of constructing unphysical
exact simulated data is to compare the performance of our new method with the SC approach.
Indeed, knowing the exact cosmology permits to compare the bias of the estimators for the statefinders in both
approaches with higher precision.

Our results indicate that the width of the posterior distributions are similar to those of dispersed simulated data,
with \emph{Eis} providing smaller standard deviations than SC.
Keeping this in mind, here we focus our attention to the statefinders' bias.

In Table \ref{Table:BiasExact_j0s0} we report the bias in $j_0$ and $s_0$, we do not show the bias on $q_0$, which is similar for both methods
and less than $2\%$. In Table \ref{Table:BiasExact_Dchi2} we show the bias statistics $\Delta \chi^2$, noting again that the values for both
methods are comparable, although \Eis has a smaller FoM.

We note from Table \ref{Table:BiasExact_Dchi2} that there is a clear tendency to have larger bias for larger matter abundances for the \emph{Eis} method.
This could have been foreseen due to the fact that modulus distance curves $\mu(z;\Omega_m)$ are more densely distributed for larger values of $\Omega_m$, and
consequently the cosmographic parameters are more sensitive to small steps on larger matter abundances. The tendency is not clear for the case of SC.

\begingroup
\squeezetable
\begin{table*}
\caption{Bias for statefinders parameters $j_0$ and $s_0$. Using exact simulated data supernovae catalogs. We show it for
SC and \emph{Eis}. }
\begin{tabular}{@{}ccccccccccccccccccccccccc@{}}
 \toprule
& \\[-6pt]
& &&&& && \multicolumn{5}{c}{$j_0$ bias} && && && \multicolumn{5}{c}{$s_0$ bias}\\
\cmidrule{5-15} \cmidrule{15-25}
& &&&& \multicolumn{9}{c}{$\Omega_m$} && \multicolumn{9}{c}{$\Omega_m$}\\
\cmidrule{6-14} \cmidrule{16-24}
\phantom{a} & $w$   &&&\phantom{ab}    & 0.25 &\phantom{a}& 0.28 &\phantom{a}& 0.30 &\phantom{a}& 0.32 &\phantom{a}& 0.35
                    &\phantom{abcdef} & 0.25 &\phantom{a}& 0.28 &\phantom{a}& 0.30 &\phantom{a}& 0.32 &\phantom{a}& 0.35  &\phantom{a}   \\
\midrule
\emph{Eis}:\\
 &-0.90 &&&& 0.053 && 0.019  && -0.008 && -0.014 && -0.054 &&  0.227 && 0.135 && 0.069  &&  0.027  && -0.028 &\\

 &-0.95 &&&& 0.026 && -0.007 && -0.033 && -0.069 && -0.099 &&  0.221 && 0.083 && 0.001  && -0.059 && -0.158 \\
		
 &-1    &&&& 0.076 && -0.011 && -0.048 && -0.083 && -0.158 &&  0.345 && 0.077 && -0.038 && -0.136 && -0.285 \\

 &-1.05 &&&& 0.098 && 0.001  && -0.063 && -0.127 && -0.190 &&  0.456 && 0.135 && -0.064 && -0.240 && -0.407 \\

 &-1.10 &&&& 0.073 && 0.020  && -0.059 && -0.171 && -0.186 &&  0.526 && 0.266 && -0.003 && -0.320 && -0.472 \\[4pt]

SC:\\
&-0.90 &&&& 0.165 && 0.090 && 0.084 && 0.010 && 0.005 &&       1.925 && 1.601 && 1.582 && 1.307 && 1.125 &\\
&-0.95 &&&& 0.187 && 0.144 && 0.075 && 0.097 && 0.046 &&       2.546 && 2.134 && 1.838 && 1.781 && 1.487 &\\
&-1    &&&& 0.255 && 0.195 && 0.148 && 0.144 && 0.075 &&       3.067 && 2.720 && 2.437 && 2.288 && 1.927 &\\
&-1.05 &&&& 0.303 && 0.257 && 0.222 && 0.090 && 0.108 &&       3.736 && 3.483 && 3.205 && 2.473 && 2.417 &\\
&-1.10 &&&& 0.316 && 0.299 && 0.261 && 0.203 && 0.145 &&       4.310 && 4.170 && 3.802 && 3.402 && 2.958 &\\
\bottomrule
\end{tabular}
\label{Table:BiasExact_j0s0}
\end{table*}
\endgroup

\begingroup
\squeezetable
\begin{table*}
\caption{Bias statistics $\Delta \chi^2$. Using exact simulated data supernovae catalogs. We show it for
SC and Eis methods.}
\begin{tabular}{@{}ccccccccccccccccccccccccc@{}}
 \toprule
& \\[-6pt]
& &&&& && \multicolumn{5}{c}{\emph{Eis}} && && && \multicolumn{5}{c}{SC}\\
\cmidrule{5-15} \cmidrule{15-25}
& &&&& \multicolumn{9}{c}{$\Omega_m$} && \multicolumn{9}{c}{$\Omega_m$}\\
\cmidrule{6-14} \cmidrule{16-24}
\phantom{a} & $w$   &&&\phantom{ab}    & 0.25 &\phantom{a}& 0.28 &\phantom{a}& 0.30 &\phantom{a}& 0.32 &\phantom{a}& 0.35
                    &\phantom{abcdef} & 0.25 &\phantom{a}& 0.28 &\phantom{a}& 0.30 &\phantom{a}& 0.32 &\phantom{a}& 0.35  &\phantom{a}   \\
\midrule

 &-0.90 &&&& 0.164 && 0.250  && 0.493  && 0.619  && 1.560  &&  0.165 && 0.127  && 0.096  && 0.068  &&  0.034 \\

 &-0.95 &&&& 0.250 && 0.441  && 0.966  && 1.655  && 3.594  &&  0.350 && 0.398  && 0.378  && 0.289  && 0.198\\
		
 &-1    &&&& 0.207 && 0.624  && 1.101  && 1.750  && 5.239  &&  0.934 && 0.913  && 0.809  && 0.790  && 0.534\\

 &-1.05 &&&& 0.221 && 0.541  && 1.142  && 2.580  && 5.652  &&  1.798 && 1.696  && 1.558  && 1.539  && 1.075\\

 &-1.10 &&&& 0.185 && 0.427  && 0.826  && 2.389  && 4.002  &&  3.371 && 2.815  && 3.036  && 2.823  && 2.483\\[4pt]

\bottomrule
\end{tabular}
\label{Table:BiasExact_Dchi2}
\end{table*}
\endgroup

To end up this section, with the $\Omega_m=0.3$, $w=-1$ exact simulated catalog, we perform a test that lets us observe the consequences of the bias
in a simple way and by one single parameter. The test consists in fixing the \eis to the $\Lambda$CDM model. Hence we
set $j_0=1$ and $s_0 = -2 - 3 q_0$, that can be translated to the \eis parameters as $E_2 =  - E_1^2 + 2 E_1$ and
$E_3 = 3 E_1^3-6 E_1^2+2 E_1$. We impose these conditions internally in the code, and let $E_1$ to be the only variable.
In Fig.~\ref{fig:fixtolcdm}, we show the posteriors of $\Omega_m$ for this test. The obtained means and standard deviations are
\begin{align} \label{1dtest}
 \Omega_m|_{\Lambda \text{CDM}}               &= 0.299 \pm 0.012, \nonumber\\
 \Omega_m|_{\text{\emph{Eis}}}                &= 0.297 \pm 0.012, \nonumber\\
 \Omega_m|_{\text{SC}}                        &= 0.275 \pm 0.011, \\
 \Omega_m|_{\text{\emph{hybrid\_1}}}          &= 0.307 \pm 0.013, \nonumber\\
 \Omega_m|_{\text{\emph{hybrid\_2}}}          &= 0.301 \pm 0.011. \nonumber
\end{align}
The dispersions are similar for the five models. Albeit the best fit is quite different for SC.
This is a consequence of the large bias of $j_0$ and $s_0$ that now has been absorbed by $q_0$.
Given that we are fitting to exact simulated data, this bias is intrinsic to the cosmographic method; a reduction of the error bars or, equivalently, adding
more data over the same redshift domain, reduces the standard deviations of the estimations, but it has a small impact on the bias.
Therefore, this one-parameter diagnostic provides a robust criterion to discard non-viable cosmography approaches.
The methods proposed in this work are not free of this bias propagation effect, although it is much smaller.

\subsection{Redshift distributions}\label{Subsec:RSdist}

We now explore different redshift distributions.
This analysis shows the importance of having a large amount of low redshift supernovae to reduce the bias.
We split the redshift range covered by the JLA catalog ($z \in (0.01,1.3)$) in four bins divided by
the redshift cuts $z_{low}$, $z_{mid}$ $z_{high}$ as in Eq.~(\ref{redshiftscuts}).
We explore three different distributions: \emph{zdist\_1} has the same redshifts as the JLA compilation,
%\emph{zdist\_4} is an even distribution over the whole range
%spanned by JLA,
\emph{zdist\_2} is skewed to low $z$, and \emph{zdist\_3} is skewed to the high $z$. The number of supernovae per bin is denoted by
$(N_1,N_2,N_3,N_4)$ with $N_1$ the number of supernovae in the bin $0<z<z_{low}$, $N_2$ the number of supernovae in the bin $z_{low}<z<z_{mid}$,
$N_3$ the number of supernovae in the bin $z_{mid}<z<z_{high}$, and $N_4$ the number of supernovae in the bin $z > z_{high}$.
For \emph{zdist\_2} and \emph{zdist\_3} the supernovae are evenly distributed over each bin.
%and analogously for the other three bins
The three distributions are binned as
\begin{align} \label{zdists}
 \text{\emph{zdist\_1}} &: \quad (111,414,181,34), \nonumber\\
% \text{\emph{zdist\_4}} &\quad (23,201,287,229) \\
 \text{\emph{zdist\_2}} &:\quad (400,200,100,40), \\
 \text{\emph{zdist\_3}} &:\quad (10,90,200,440). \nonumber
\end{align}
With these distributions we construct exact simulated data based on a $\Lambda$CDM model with $\Omega_m=0.3$ and
errors $\sigma_\mu = 0.118$, which is the average of errors in the JLA catalog.

In Table \ref{Table:zdist} we show the marginalized 1-dimensional estimations, along with the 3-dimensional bias statistics $\Delta \chi^2$ and FoM.
It is expected that the standard deviations of the higher statefinders are smaller when we add more supernovae at high redshifts, because
the influence of variations in $\Omega_m$ on $d_L$ is more evident at higher redshifts.
Nevertheless it is clear that the bias is enhanced for the \emph{zdist\_3} case, having it higher FoM and $\Delta \chi^2$. Contrary,
when considering larger amounts of low redshifts supernovae, the bias is considerably reduced, this is the case of \emph{zdist\_1}.

\begingroup
\squeezetable
\begin{table*}
\caption{Different z distributions for the \emph{Eis} method. The distributions are defined in Eq.~(\ref{zdists}).}
\begin{tabular}{@{}cccccccccccccccccc@{}}
 \toprule
 \\[-2pt]
\phantom{a} & Redshift Distribution   &&& \phantom{abc}    & $\hat{q}$ &\phantom{abc}& $\hat{j}$ &\phantom{abc}& $\hat{s}$
&\phantom{abc}& $\Delta \chi^2$ &\phantom{abc}& FoM
&\phantom{abcd}& $\hat{\Omega}_M$ (1 parameter test) \\[4pt]
\midrule\\

 &\emph{zdist\_1} &&&& $-0.55 {}^{+0.08}_{-0.07}$ && $ 0.94 {}^{+0.29}_{-0.48}$  && $-0.42 {}^{+0.10}_{-0.43}$
                  && $0.571$ && $0.010$           && $ 0.299 \pm 0.012$ \\[7pt]

% &\emph{zdist\_2} &&&& $-0.54 \pm 0.09$           && $ 0.86 {}^{+0.29}_{-0.43}$  && $-0.45 {}^{+0.05}_{-0.32}$
%                  && $1.623$ && $0.007$           && $ 0.275 \pm 0.010 $ \\[7pt]
		
 &\emph{zdist\_2} &&&& $-0.55 {}^{+0.08}_{-0.07}$ && $ 0.93 {}^{+0.30}_{-0.49}$  && $-0.45 {}^{+0.07}_{-0.37}$
                  && $0.816$ && $0.009$           && $ 0.300 \pm 0.012$ \\[7pt]

 &\emph{zdist\_3} &&&& $-0.53 {}^{+0.10}_{-0.11}$ && $ 0.84 {}^{+0.35}_{-0.47}$  && $-0.38 {}^{+0.08}_{-0.42}$
                  && $1.003$ && $0.013$           && $ 0.259 {}^{+0.010}_{-0.009} $ \\[4pt]

\bottomrule
\end{tabular}
\label{Table:zdist}
\end{table*}
\endgroup

This is more evident if we use the one-parameter diagnostic of Sec.~\ref{subsec:exactSD}. In Fig.~\ref{fig:fixtolcdmZBins}
we show the plots for $\Omega_m$ obtained by performing this test and in Table \ref{Table:zdist} we show the
best fits and 0.68 \emph{c.i.}. It is clear that the inclusion of a large amount of low
redshift supernovae in \emph{zdist\_1} reduces the bias, while for \emph{zdist\_3} it becomes quite large, such that the
true value is not attainable. Instead, the standard deviation is reduced as explained above.

\begin{figure}
\begin{center}
\includegraphics[width=2.5in]{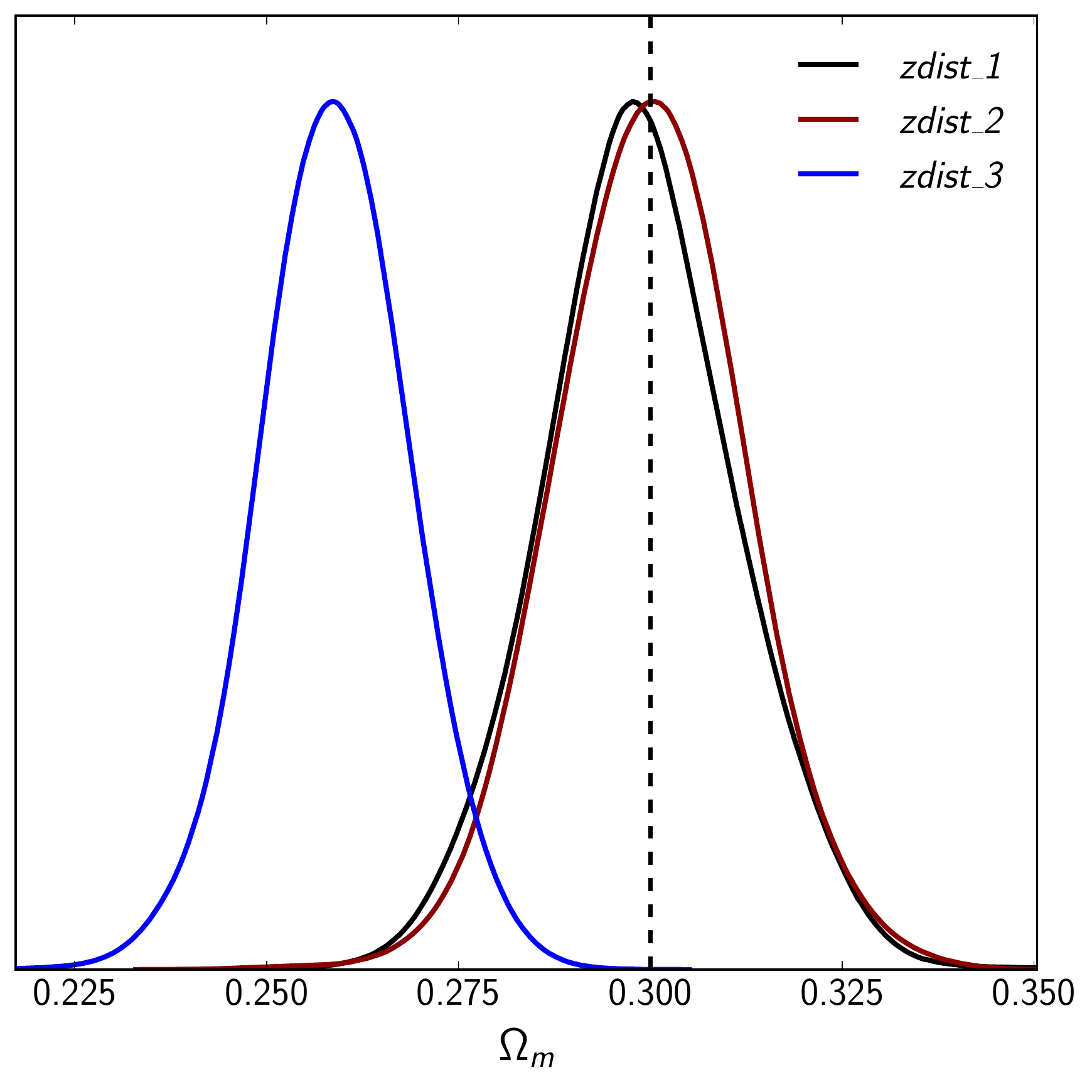}
\caption{One-parameter diagnostic. We show the estimated $\Omega_m$ for \emph{zdist\_1} (black curve), \emph{zdist\_2} (red curve) a
nd \emph{zdist\_3} (blue curve) as defined in Eq.~(\ref{zdists}). We let
vary $E_1$  and fix the rest \emph{eis} parameters with the degeneracies in $\Lambda$CDM. The matter abundance is obtained as
$\Omega_m= \frac{3}{2}E_1$. As a reference, we show the true value $\Omega_m=0.3$ (dashed vertical line).}
\label{fig:fixtolcdmZBins}
\end{center}
\end{figure}

% The reason behind this, is that
% the method is giving poor convergence at high redshifts, we show in fig. X plots for $d_L$ obtained by setting the \eis
% parameters to $\Lambda$CDM with $\Omega_m =0.30$

\end{subsection}

\end{section}

\begin{section}{Fitting the statefinders to the JLA and Union 2.1 compilations} \label{sec:rdanalysis}

In this section we fit the \emph{Eis} and SC methods to real data. To this end we consider the two most used supernova type Ia compilations:
the JLA \cite{Betoule:2014frx} and the Union2.1 \cite{Suzuki:2011hu}.

The major complication on real supernova catalogs is the presence of systematic errors,
mainly due to photometric calibration and selection bias, but also to physical effects as photon absortion by intervening dust and gravitational
lensing. Supernova systematics specially limits the fitting procedure when considering several parameters, as it is our case. Since these
errors do not follow any specific physical model (\emph{e.g.} $\Lambda$CDM) we expect departures on the bias obtained in the previous sections,
specially for the parameter $j_0$, which serves as a null-diagnostic for flat $\Lambda$CDM
---$j_0$ will not be necessarily underestimated (overestimated) for the \Eis (SC) method,
as it is the case for the simulated data. To partially alleviate these problems we could additionally use other types of data, as BAO or direct observations
of the Hubble flow. Although straightforward, we delegate this endeavor for future investigations, since in this work we have only
analyzed the response of the statefinders to supernovae data.

The numerical fit to JLA is quite slow due to the incorporation of nuisance parameters that are present also on the covariance error matrix. Thus, for
every step on the Markov chains one has to invert the full covariance matrix. In the presence of nuisance parameters, the modulus distance to
the $k$-\emph{th} supernova on the compilation follows the
linear model $\mu_k = m_{B,\,k} - (M_{B,\,k} -\alpha_{JLA} X_{1,\,k}+ \beta_{JLA} C_k)$ where $m_B$ is the observed peak magnitude, $M_B$ the absolute magnitude,
$X_1$ the time stretching of the light curve, and $C$ the color at maximum brightness.
In principle the absolute distance is the sum of two nuisance parameters $M_B = M_B^1 + \Delta_M$, but since the errors do not depend on them we can
internally marginalize, as it is already incorporated in the JLA likelihood module to CosmoMC.
Therefore, we only estimate $\alpha_{JLA}$ and $\beta_{JLA}$ nuisance parameters.

We were unable to sample adequately the tails of the posterior distributions for $E_3$ (or equivalently the snap) in the SC case using the JLA compilation.
For that reason we warn the reader to take the JLA results for SC in Table \ref{Table:fitstorealdata} with precaution.
We also report the $\chi^2$ statistics maximum values; the number of degrees of freedom (d.o.f.) for this compilation considering three free parameters and
two nuisances is $\text{d.o.f.} = 740 - 3 -2 = 735$.

The nuisance parameters for both SC and \emph{Eis} have almost the same 0.68 \emph{c.i.}: $\alpha_{JLA} = 0.14 \pm 0.01$, $\beta_{JLA} = 3.11 \pm 0.08$,
differing only in the third significative figure.

We perform a second fit to the \emph{Eis} method, but with the nuisance fixed to their best fit values.
We name this fit as \emph{Eis*}.

The Union2.1 compilation is lacking of nuisance parameters, thus the fitting procedure is straightforward.
Considering systematic errors the statefinders constraints at 0.68 \emph{c.i.} are shown in
Table \ref{Table:fitstorealdata}. Additionally, we report the $\chi^2$ statistics maximum value, which for this compilation and
for three free parameters has $\text{d.o.f.} =580 - 3 = 577$.

Again, for the case of SC we obtained non-conclusive posterior distributions for $E_3$. We conclude that the available supernova data compilations
alone are not able to fit SC when using more than two statefinder parameters.

In Table \ref{Table:fitstorealdata} the central values are the means of the marginalized posterior distributions, and the errors denote the
departures from the means of the lower and higher limits of the 0.68 \emph{c.i}. For the \emph{Eis} method fitting to the Union 2.1 compilation,
due to the skewness of $s_0$, its mean value is located out of its 0.68 \emph{c.i}

\begingroup
\squeezetable
\begin{table*}
\caption{Estimations for \emph{eis} and statefinder parameters to JLA (top) and Union2.1 (bottom) compilations.
\emph{Eis*} fixes the nuisances to $\alpha_{JLA} = 0.141$, $\beta_{JLA} = 3.108$. }
\begin{tabular}{@{}llccccccccccccccc@{}}
\toprule
&\\[-3pt]
             && \phantom{abcd} $E_1$\phantom{abcd} & \phantom{abcd} $E_3$ \phantom{abcd} & \phantom{abcd} $E_3$ \phantom{abcd} &\phantom{abc}&
                \phantom{abcd} $q_0$\phantom{abcd} & \phantom{abcd} $j_0$ \phantom{abcd} & \phantom{abcd} $s_0$ \phantom{abcd} &&
             &&& $\chi^2$ & \\

&\\[-3pt]
\midrule \\[-3pt]

\underline{JLA} \\[5pt]

\emph{Eis} &&  $ 0.45^{+0.15}_{-0.12}$ & $ 0.76^{+0.37}_{-0.81}$ & $-0.60^{+1.00}_{-0.30}$ &&
               $-0.54^{+0.15}_{-0.12}$ & $ 1.07^{+0.43}_{-0.98}$ & $ 0.50^{+0.19}_{-1.30}$ &&
                & && 705.4 &\\[7pt]

\emph{Eis*} &&  $ 0.43^{+0.10}_{-0.08}$ & $ 0.86^{+0.25}_{-0.50}$ & $-0.48^{+0.61}_{-0.13}$ &&
                $-0.57^{+0.10}_{-0.08}$ & $ 1.20^{+0.31}_{-0.63}$ & $ 0.23^{+0.02}_{-0.76}$ &&
                &  && 820.5 &\\[7pt]

SC          && $ 0.49^{+0.16}_{-0.23}$ & $ 0.85^{+1.96}_{-1.33}$ & $4.01^{+8.71}_{-4.12}$ &&
               $-0.51^{+0.16}_{-0.23}$ & $ 1.15^{+2.12}_{-1.19}$ & $4.58^{+2.58}_{-7.54}$ &&
                &  && 698.7 &\\[9pt]

\underline{Union2.1} \\[5pt]

\emph{Eis} &&  $0.36^{+0.15}_{-0.13}$ & $0.99^{+0.47}_{-0.98}$ & $-0.64^{+1.30}_{-0.05}$ &&
               $-0.64^{+0.15}_{-0.13}$ & $1.42^{+0.58}_{-1.2}$ & $0.95^{-0.06}_{-1.80}$ &&
                &  && 568.7 &\\[7pt]

SC         &&  $0.39^{+0.15}_{-0.18} $ & $1.08^{+1.36}_{-1.41}$ & $-3.94^{+8.42}_{-3.68}$ &&
               $-0.61^{+0.15}_{-0.18}$ & $1.49^{+1.52}_{-1.61}$ & $4.44^{+2.69}_{-7.85} $ &&
                &  && 572.6 &\\[7pt]

\bottomrule
\end{tabular}
\label{Table:fitstorealdata}
\end{table*}
\endgroup

\end{section}

\begin{section}{Conclusions}  \label{sec:concl}

In this work, by accounting for
Hubble expansions inside the luminosity distance, we tailored a method that leads to less biased estimations and smaller standard deviations than those
in SC. We baptized
our new method \emph{Eis}, since it estimates the coefficients (named
\emph{eis}) of Taylor's expansion of the normalized Hubble rate
$E(z)=H(z)/H_0$.
We focused on the first three \emph{eis} parameters by using data spanning over the redshift interval $0<z \leq 1.2$. From them, the
deceleration $q_0$, jerk $j_0$, and snap $s_0$ statefinders parameters can be derived.

In order to speed up the computations, we further split all the numerical
analyses in redshift bins.
In so doing, the order of Hubble's expansion
depends on the
redshift of a single supernova data. We have checked that this
binning procedure does
not alter significatively the parameters' estimations.
This speeding up is in average a factor of 2.5, and it is mainly due to the convergence
of MCMC chains, which is attained within a less number of steps.

% We gave, as theoretical support to our formalism, a brief discussion in
% which we proved that pure Taylor expansions \tcomm{Do you mean direct Taylor expansion to the luminosity distance? If so, you can say it.
% Otherwise, this has no meaning.}
% would less rapidly converge to
% approximate the exact luminosity distance definition \tcomm{less than what? }.

We further constructed two hybrid models that made use of SC
for redshifts $z<z_{mid}=0.4$, whereas over the rest of the redshift domain
they employed expansions of our \emph{Eis} method.

By extensively considering simulated catalogs of supernovae built up through
the JLA compilation, we also compute several bias statistics for cosmological
models near the concordance model. Specifically, for each single parameter,
we compute the bias and the risk statistics; and,
in order to account for the correlations of the statefinders, we further use the
$\Delta \chi^2$ bias statistics and the FoM.

We concluded that our methods provide less biased estimations than in the
SC case. Moreover, the standard deviations of the posterior distributions
are considerably smaller than in SC.

The other issue that we faced out was the one due to degeneracies. We
showed that SC is not able to follow the $\Lambda$CDM degeneracies on the
statefinders, even when the fitting is performed against an ``exact'' $\Lambda$CDM model simulated catalog. Meanwhile
our method \emph{Eis} is capable to do it.
We assumed that the third-degree polynomial form of the $z^4$ coefficient in the
SC luminosity distance is responsible for these fictitious degeneracies.
Actually, this intuition motivated us in the first place to
construct more viable methods in cosmography.

We further proposed a new test to reject non-viable methods in cosmography.
It consists in building up an exact simulated catalog of supernovae by
means of a given fiducial $\Lambda$CDM cosmology.
Thereafter, all except the $E_1$ (or equivalently $q_0$) of the
cosmographic parameters should be fixed into the code
by using the degeneracies of the $\Lambda$CDM model. Finally, the estimations are performed.
This is a simple 1-parameter diagnostic that
addresses the propagation of bias from higher orders statefinders to the
deceleration parameter. We showed how SC is not able to pass this test since the
estimation of $\Omega_m$ for it turns out to be more than $2\,\sigma$ away from
the true value, reflecting its highly biased estimations.
We highlighted that the methods presented here are not completely free of this bias propagation, albeit the effects are much smaller
than in SC.

Finally, we applied our method to real data, which provides the most stringent to date constraints of the statefinders by using supernovae data only; for either
JLA or Union2.1 compilations.

We make publicly-available a module to the code CosmoMC that
perform the MCMC numerical analysis for the cosmographic methods of this
work at \href{https://github.com/alejandroaviles/EisCosmography}{https://github.com/alejandroaviles/EisCosmography}.
There, we also uploaded all the simulated data as well as further tables and statistical files.

Future works will focus on investigating the same procedures, discussed in our work, for investigating bias and dispersions of cosmographic estimations for other kinds of surveys such as BAO and/or $H(z)$ data. For $H(z)$ data, in particular, we expect the results to be the same for both {\emph Eis} and SC methods. This consideration comes from the definitions of both the methods. In fact, when assuming $H(z)$ data,
SC needs the direct expansions of the Hubble rate. This fact turns out to give analogous outcomes with respect to our approach.

\end{section}

\acknowledgements
We would like to thank Prof. P.K.S. Dunsby for reasonable discussions.
This work was partially supported
by ABACUS, CONACyT grant EDOMEX-2011-C01-165873, and
National Research Foundation (NRF). A.A. wants to thank the CONACyT Fronteras Project 281.

% \begin{widetext}
\appendix

\begin{section}{Useful relations of cosmography} \label{app::formulas}

Cosmography attempts to consider the fewest number of assumptions as possible. Its basic demand is that the background cosmology
is well described by a FRW universe, at least at very large scales. Lying on this assumption, it expands the scale factor $a(t)$ in Taylor series
about an arbitrary cosmic time $t_*$. That is,
\begin{eqnarray}\label{serie1a}
a(t) &=& \sum_{n=0}^\infty \frac{1}{n!} \frac{d^n a (t) }{dt^n}\Big|_{t=t_*}  \Delta t^n \nonumber\\
&=& a_* \sum_{n=0}^\infty \frac{1}{n!} \frac{1}{a_* H_*^n} \frac{d^n a (t) }{dt^n}\Big|_{t=t_*}  (H_* \Delta t)^n,
\end{eqnarray}
where $a_* = a(t_*)$, $H_*=H(t_*)$, and $\Delta t \equiv t-t_*$. From this expansion we define the hierarchy of statefinders, the first three are
given by
\begin{eqnarray}
 q &\equiv& -\frac{1}{a H^2} \frac{d^2a}{d t^2}, \nonumber\\
 j & \equiv& \frac{1}{a H^3} \frac{d^3a}{d t^3}, \\
 s &\equiv& \frac{1}{a H^4} \frac{d^4a}{d t^4}.  \nonumber
\end{eqnarray}
These definitions are the most used in the literature, but differ from the originals defined in \cite{Sahni:2002fz}.
In this work we concentrate only in the statefinders at present
time, that is in $q_0$, $j_0$, and $s_0$.

By inverting  Eqs.~(\ref{eisOfsf}) we get the statefinders in terms of the \emph{eis} parameters.
\begin{eqnarray}
q_0 &=& -1 + E_1,  \nonumber\\
j_0 &=& 1 - 2 E_1 + E_1^2 + E_2, \\
s_0 &=& 1 - 3 E_1 + 3 E_1^2 - E_1^3 + E_2 - 4 E_1 E_2 - E_3. \nonumber
\end{eqnarray}
Then, we can substitute in Eq.~(\ref{dL_sc}) to get the luminosity distance of standard cosmography
as a function of the \emph{eis} parameters
\begin{eqnarray} \label{dlSCeis}
 & & \tilde{d}_{L \text{(SC)}}(z;E_1,E_2,E_3) = z + \frac{1}{2}(2 - E_1)z^2 \nonumber\\
 & & + \frac{1}{6}(-3 E_1 + 2 E_1^2 - E_2)z^3  \nonumber\\
 & &   + \frac{1}{24}(8 E_1^2 - 6 E_1^3 - 4 E_2 + 6 E_1 E_2 - E_3) z^4, \nonumber
\end{eqnarray}
which is the expression we use in our numerical computations.

The hybrid methods consist in using SC with two parameters up to the redshift $z_{mid}$, and beyond it expand the integrand in $d^{(n)}_L$. Specifically,
the luminosity distance for \emph{hybrid\_1} method is

\begin{widetext}

\begin{equation} \label{hybridsModel}
   \tilde{d}_{L\text{(\emph{hybrid\_1})}}(z;E_1,E_2,E_3) = \begin{cases}
                               & \\[-8pt]
               \tilde{d}_{L \text{(SC)}}(z;E_1)               & \qquad z \leq z_{low}\\
                               & \\[-8pt]
               \tilde{d}_{L \text{(SC)}}(z;E_1,E_2)               & \qquad z_{low} < z \leq z_{mid}\\
                               & \\[-8pt]
               (1+z)\Big( z + \frac{1}{2}E_1 z^2 + \frac{1}{6} (2 E_1^2 + E_2) z^3  & \\
              \qquad \qquad+ \frac{1}{24} (-6 E_1^3 + 6 E_1 E_2 -E_3) z^4 \Big)     & \qquad z_{mid}< z \leq z_{high}\\
                               & \\[-8pt]
               (1+z)\Big( z + \frac{1}{2}E_1 z^2 + \frac{1}{6} (2 E_1^2 + E_2) z^3 &  \\
                     \qquad  \qquad+ \frac{1}{24} (-6 E_1^3 + 6 E_1 E_2 -E_3) z^4     &  \\
\qquad \qquad+ \frac{1}{60} (12 E_1^4 - 18 E_1^2 + 3 E_2^2 + 4 E_1 E_3) z^5 \Big)     & \qquad \, z > z_{high}

           \end{cases}
\end{equation}

The luminosity distance for \emph{hybrid\_2} method is the same as that for \emph{hybrid\_1} but with  $z_{high} = z_{mid} = 0.4$.

\end{widetext}

Other useful formulae are the relations between flat $w$CDM and the cosmographic parameters. These are given by
% \begin{equation}
%  q_{0,\Lambda\text{CDM}} = -1 + \frac{3}{2} \Omega_m, \qquad j_{0,\Lambda\text{CDM}} = 1, \qquad  s_{0,\Lambda\text{CDM}}= 1- \frac{9}{2} \Omega_m.
% \end{equation}
%
%
% \begin{equation}
%  E_{1,\Lambda\text{CDM}} =  \frac{3}{2} \Omega_m, \qquad E_{2,\Lambda\text{CDM}} = 3 \Omega_m - \frac{9}{4} \Omega_m^2,
%  \qquad  E_{3,\Lambda\text{CDM}}= 3 \Omega_m - \frac{27}{2} \Omega_m^2 + \frac{81}{8} \Omega_m^3.
% \end{equation}
\begin{eqnarray} \label{app:sflcdm}
 q_{0,w\text{CDM}} &=& \frac{1}{2} + \frac{3}{2} w (1- \Omega_m), \nonumber\\
 j_{0,w\text{CDM}} &=& 1 + \frac{9}{2}w(1+w)(1-\Omega_m), \\
 s_{0,w\text{CDM}} &=& - \frac{7}{2} - \frac{81}{4} w (1-\Omega_m) - \frac{9}{4}w^2(16 - 19\Omega_m + 3\Omega_m^2) \nonumber\\
                   & & - \frac{27}{4}w^3(3- 4 \Omega_m + \Omega_m^2 ), \nonumber
\end{eqnarray}
and
\begin{eqnarray}
 E_{1,w\text{CDM}} &=& \frac{3}{2} + \frac{3}{2} w (1- \Omega_m), \nonumber\\
 E_{2,w\text{CDM}} &=& \frac{3}{4} + 3 w (1- \Omega_m) + \frac{9}{4} w^2 (1-\Omega_m^2), \\
 E_{3,w\text{CDM}} &=& -\frac{3}{8} - \frac{3}{8}w (1-\Omega_m) + \frac{27}{8}w^2(1-\Omega_m^2)  \nonumber\\
                   & & + \frac{27}{8}w^3(1- \Omega_m + 3 \Omega_m^2  -  3 \Omega_m^3). \nonumber
\end{eqnarray}

\end{section}

% \end{widetext}
\begin{section}{Binning Analysis in the \emph{Eis} method}

Throughout this work we have assumed that a suitable choice for the redshift cuts is given by Eq.~(\ref{redshiftscuts}). We
now check the goodness of this approach. In particular, we do not perform the analysis for the hybrid methods
because the \emph{Eis} paradigm works better beyond the 1-parameter test and turns out to be more appropriate.

First, we note that $z_{low}$ and $z_{mid}$ cuts are physically different than
$z_{high}$. Indeed, the three aforementioned bins make use of  Eqs.~(\ref{EisModel}) for the estimation of
the \emph{eis} parameters, whereas for $z > z_{high}$ we require the integrand Taylor expansion.

We highly emphasize that the choice of $z_{low}$ and $z_{mid}$ is tailored only for speeding up the numerical outcomes, and due to this fact  we decide to test three models: the first with the same binning already discussed throughout the work,
the second model without bounds over $z_{low}$ and $z_{mid}$, implying the use of $\tilde{d}_L(z) = \tilde{d}^{(3)}_L(z)$ for $z< z_{high}$ while finally the third model with nuisance parameters for the redshift cuts with uniform priors over the intervals
$0< z_{low, N}< 0.2$ and $0.3<z_{mid, N}<0.7$ (black curves).
The three methods are compared with the $\Lambda$CDM model with $\Omega_m=0.30$ by using an exact simulated catalog; we even employ the
redshift distribution found in the JLA catalog, with errors $\sigma_\mu=0.118$, as in \emph{zdist\_1} of Sec.~\ref{Subsec:RSdist}.

\begin{figure}
\begin{center}
% % \includegraphics[width=3.3in]{figures/PlotNuisances1.pdf}/home/waco/Desktop/zcuts.pdf
\includegraphics[width=3.3in]{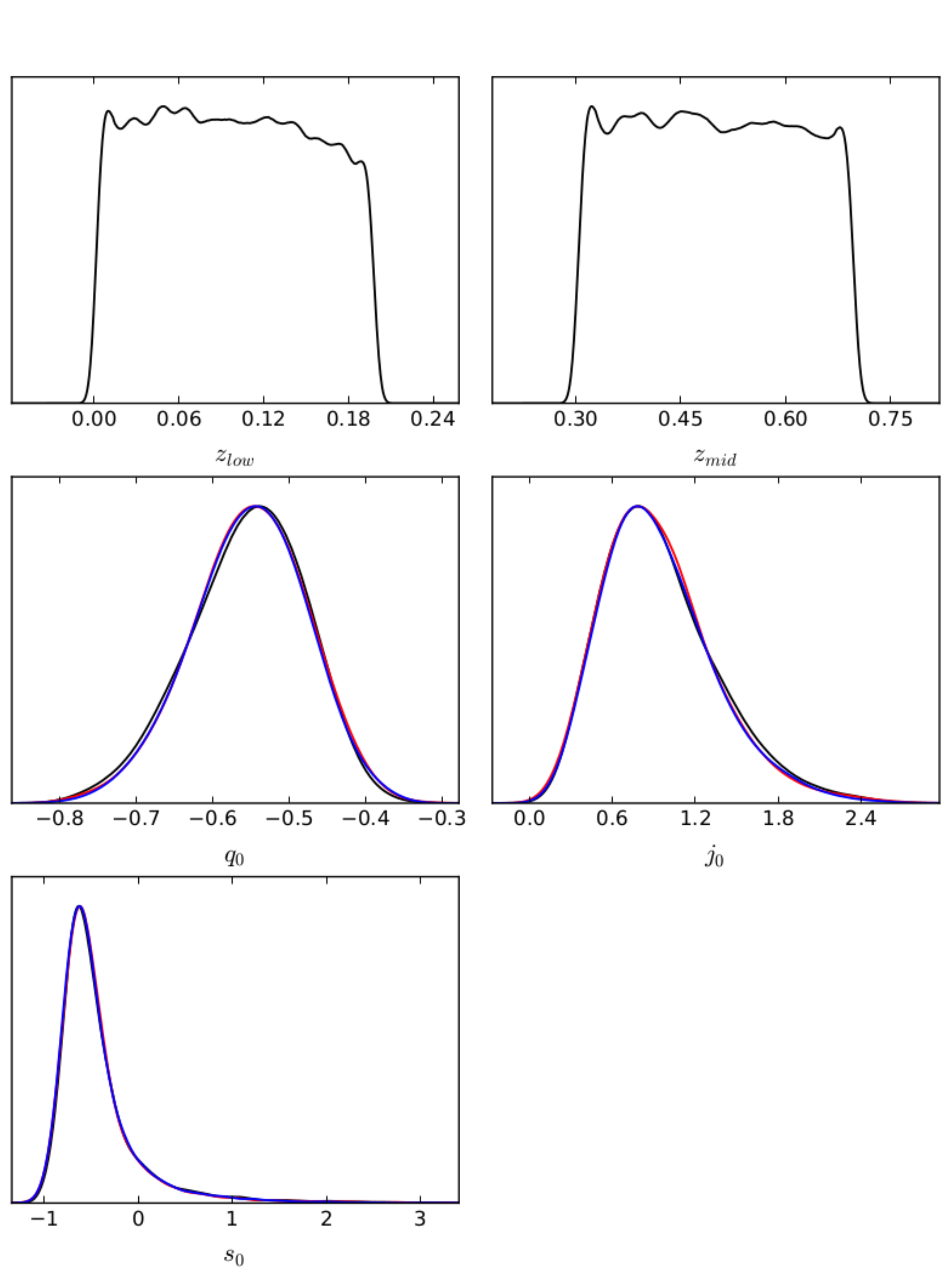}
\caption{Here we focus on the impacts of $z_{low}$ and $z_{mid}$ on our estimations. Red curves correspond to the choice of cuts used throughout this work. Blue
curves are for estimation with $z_{low} = z_{mid} = 0$. Black curves correspond to estimations using $z_{low}$ and $z_{mid}$
as nuisance parameters. We note the differences between the three models are very small.}
\label{fig:Nuisances}
\end{center}
\end{figure}

We perform MCMC estimations and we show the 1-dimensional marginalized posterior distributions in Fig.~\ref{fig:Nuisances}.
We note no differences among the estimations of each methods. Moreover, the nuisances do not show a preferred value. From this analysis we conclude
that the cuts $z_{low}$ and $z_{mid}$ are not very relevant for the estimations.

However, the situation is quite different for the fourth bin, defined by $z>z_{high}$. When it is not considered, the best fit does not differ
significatively, but the tails of the posterior distributions become very noisy, specially for $E_3$ (or equivalently $s_0$). This behavior has been
portrayed in the $q_0$-$s_0$ contour plot of Fig.~\ref{fig:highzcut} and are traced back to the samples of $E_3$ and $E_2$ that can make
the denominator of $\tilde{d}_L$ in Eq.~(\ref{EisModel}) very close to zero or even negative.

One may wonder that if considering only SNe redshifts which are smaller than the value of $z_{high}$ turns out to be more convenient for better estimations. This does not seem to be the
case if considering three parameters (although considering only two, instead of three, becomes a viable strategy).
In Fig.~\ref{fig:highzcut} we also plot the former behavior, revealing
that high redshift SNe are necessary for the estimation of the third statefinder.

\begin{figure}
\begin{center}
\includegraphics[width=3.3in]{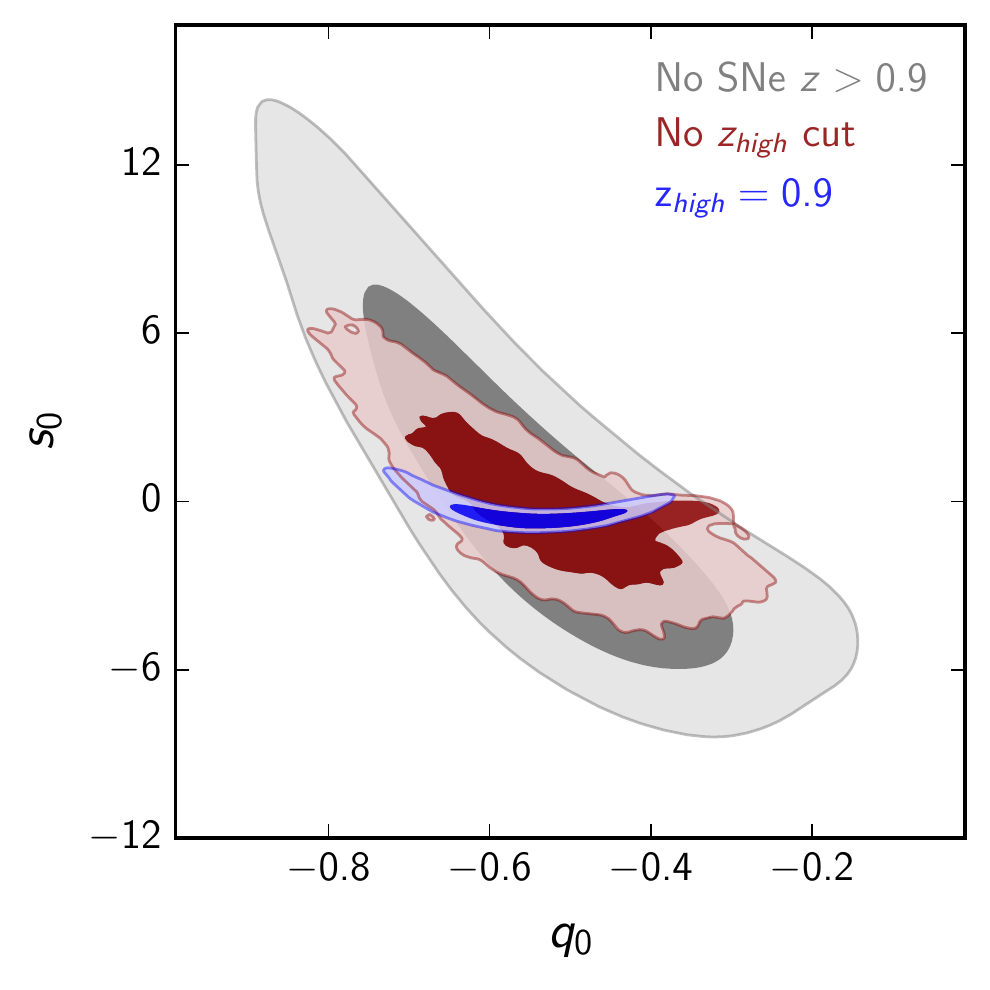}
\caption{$q_0$-$s_0$ contour plots for three models with different behaviors at high $z$. Blue contours correspond to the
standard {\emph Eis} method, the red ignores the $z_{high}$ cut, and the gray does not use SNe at redshifts higher than $z=0.9$.
This analysis forecasts the high importance of getting limits over $z_{high}$.}
\label{fig:highzcut}
\end{center}
\end{figure}

Finally, we test methods in which there is no $z_{low}$ and $z_{mid}$ cuts, and we choose $z_{high}$ to take different
values from $z=0.1$ to $z=1.3$, comprising 120 different $z_{high}$ redshifts separated by intervals of size $\Delta z =0.1$.
We perform the 1-dimensional parameter test of Sec.~\ref{subsec:exactSD}
against the same simulated catalog of Figs.~\ref{fig:Nuisances} and \ref{fig:highzcut}. In Fig.~\ref{fig:zhighOm} we show the 1-dimensional
mean (solid line) and 0.68 (blue) and 0.95 (light blue) \emph{c.i.} of the $\Omega_m$ distributions for each redshift $z_{high}$; for
comparison we also show the true $\Omega_m$ value (Dashed line). Below $z_{high} \approx 0.65 $ the plots show a constant trend
both in the best fit around $\Omega_m \approx 0.312$ and in the standard deviation $\sigma_{\Omega_m} \approx 0.015$, for intermediate redshifts $0.7 \lesssim
z_{high} \lesssim 0.9$ there is a transition, and above $z_{high} \approx 0.95$ there is also a constant trend with
$\Omega_m \approx 0.294$ and $\sigma_{\Omega_m} \approx 0.011$. At our choice, used in this work, $z_{high}=0.9$, we get
$\Omega_m = 0.297 \pm 0.012$ in agreement with Eq.~(\ref{1dtest}).

We also note that the noise present in the red curves of Fig.\ref{fig:highzcut} is not displayed in the 1-dimensional test; we
assume that this is consequence of the fact that $E_3$ and $E_2$ were fixed to $E_1$ through their $\Lambda$CDM constraints.

\begin{figure}
\begin{center}
\includegraphics[width=3in]{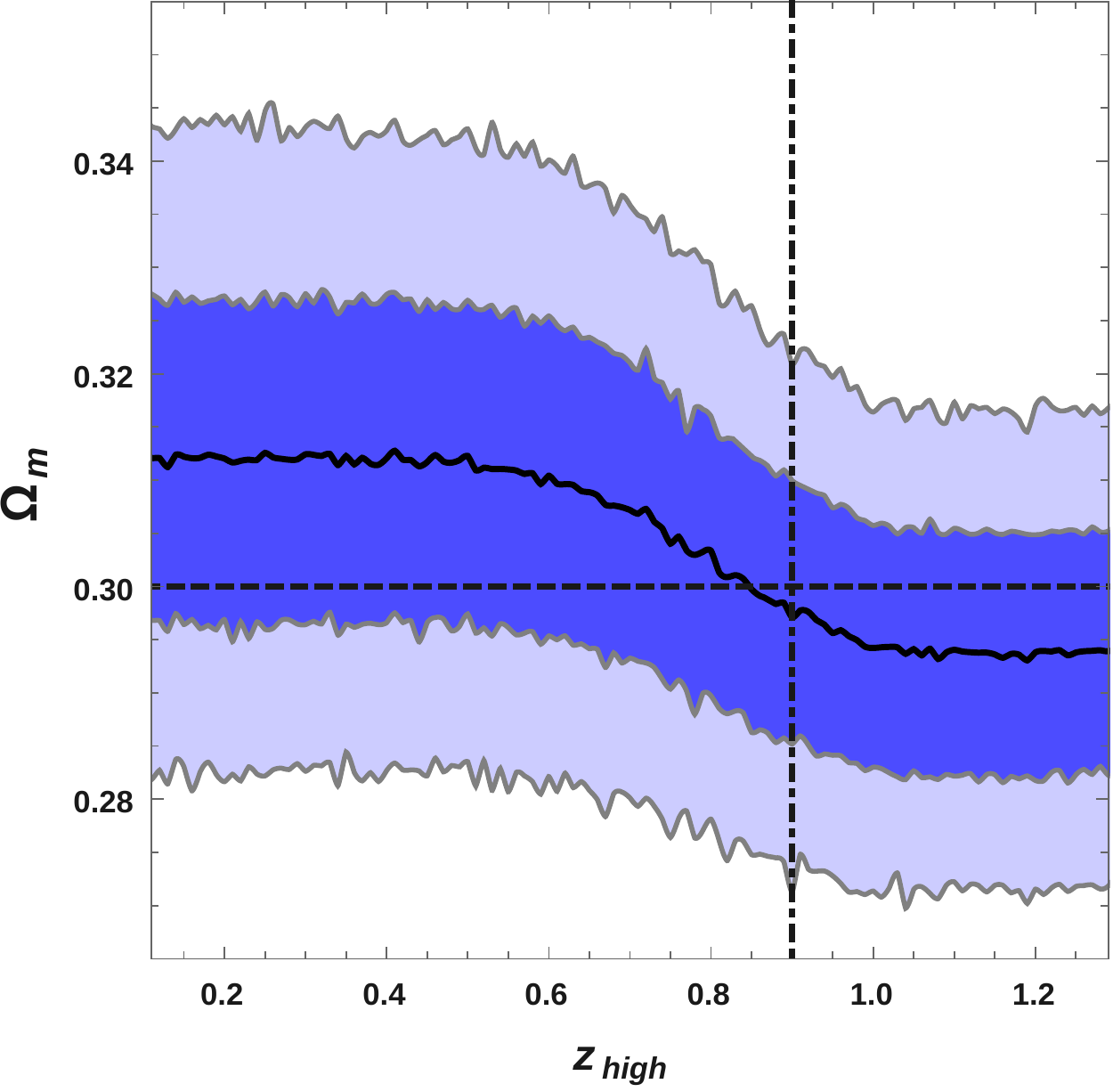}
\caption{1-dimensional estimations. We set $z_{low}=z_{mid}=0$ and let  $z_{high}$ to take different
values from $z=0.1$ to $z=1.3$. The horizontal dashed line shows the real value $\Omega_m =0.3$, and the vertical line shows the $z_{high}$ chosen throughout
this work.}
\label{fig:zhighOm}
\end{center}
\end{figure}

\end{section}

 \bibliographystyle{JHEP}  % Use the "unsrtnat" BibTeX style for formatting the Bibliography
 \bibliography{EisCG}  % The references (bibliography) information are stored in the file named "bibliography.bib"

\end{document}